\documentstyle{article}
\renewcommand{\thefootnote}{*}
\begin{document}
\newcommand{\C}{{\bf C} \!\!\!\! I}
\newcommand{\HH}{H \!\!\! I}
\newcommand{\Z}{Z \!\!\! Z}
\newcommand{\la}{\langle}
\newcommand{\ra}{\rangle}
\newtheorem{th}{Theorem}
\newtheorem{cor}{Corollary}[th]
\newtheorem{de}{Definition}
\newtheorem{pr}{Proposition}
\newtheorem{co}{Corollary}[pr]
\newtheorem{rem}{Remark}
\begin{center}
{\bf
ON AN ALGEBRAIC APPROACH TO HIGHER DIMENSIONAL}
\\
{\bf STATISTICAL MECHANICS}
\\
\vspace{.4in}
Paul  Martin
\footnote{Permanent address:
Mathematics Department,
City University,
Northampton Square, London EC1V 0HB, UK
}
and $\;$
\renewcommand{\thefootnote}{\dagger}
Hubert Saleur\footnote{On leave from SPHT CEN SACLAY 91191 Gif Sur Yvette
France}
\\
Physics Department, Yale University, New Haven CT06511  USA.
\end{center}

\vspace{.2in}

\noindent
{\bf Abstract} \hspace{.3in}
We
study
representations of Temperley-Lieb algebras associated
with the transfer matrix formulation of
statistical mechanics on arbitrary  lattices.

We first discuss  a new
hyperfinite algebra, the Diagram algebra $D_{\underline{n}}(Q)$,
 which is a
quotient of the Temperley-Lieb algebra
appropriate for Potts models in the mean field case,
and in which the algebras appropriate for
all transverse lattice shapes $G$ appear as subalgebras.
We give the complete structure of this subalgebra
 in the case ${\hat A}_n$ (Potts model on a cylinder).

The study of the Full Temperley Lieb algebra of graph $G$
reveals
 a vast number of infinite sets of inequivalent irreducible
representations characterized by one or more (complex)
parameters associated to topological effects such as links.
We give a complete classification
in the ${\hat A}_n$ case where the only such effects are loops and twists.


\section{Introduction}

Finding integrable statistical mechanics systems in dimension greater than two
is notably difficult, and very little is known about that question
\cite{all}. In two dimensions there are algebraic structures more general than
integrability, whose study nevertheless provides some physical information
\cite{Mar,others,Levy}. These  structures are not all constrained to two
dimensions.  For example the Temperley Lieb \cite{TL} algebra:
consider the complete unoriented graph of $n$ nodes,
here called $\underline{n}$, and all
those subgraphs $G \subset \underline{n}$
obtained by removing bonds (edges) from the
complete graph.
\begin{de}
We define $T_{G}(Q)$, the Full Temperley-Lieb algebra of the
graph $G$ \cite{Mar}, to be the unital algebra over $\C$
with generators
\[
<1, \; U_{i.} \;\;\;
 (i=1,2,..,n), \;  U_{ij} = U_{ji}  \;\;\; (\mbox{edge }(i,j) \in  G)>
\]
and relations:
\begin{equation}
\label{r4}
U_{}^2 = \sqrt{Q} U_{}
\end{equation}
(any indices)
\begin{equation}
\label{r5}
U_{i.} U_{ij} U_{i.} =  U_{i.}
\end{equation}
\begin{equation}
\label{r5'}
U_{ij} U_{i.} U_{ij} =  U_{ij}
\end{equation}
\begin{equation}
\label{r6}
 [ U_{i.} , U_{j.} ] =
[U_{ij},U_{kl} ] =
[U_{i.}, U_{kj} ] =0
                        \hspace{.3in} i \ne k,j.
\end{equation}
\end{de}
We note the very useful nested structure of inclusions of these algebras
(c.f. \cite{HGJ}):
\[
G \subset G' \;\; \Rightarrow \;\; T_{G}(Q) \subseteq T_{G'}(Q)
\]
where the restriction is achieved by simply omitting the
appropriate bond generators.
For example, with $G=A_n$, the $n$ node chain graph, we recover the
original Temperley-Lieb algebra $T_{2n}(Q)$.
Conversely, it
 follows  from the definition of the Potts model
\cite{Baxter}
  that $T_G(Q)$ is
a
 generalization of the
transfer matrix algebra
$T_{2n}(Q)=T_{A_n}(Q)$
appropriate for building               
a transfer matrix layer of shape $G$ \cite{Mar} -
in other words for
overall lattice shape $G \times \Z$.
This graph $G$ corresponding to the shape of physical space
is not to be confused with the
{\em configuration space} graphs of
\cite{P,PS}, which
work only for the two dimensional case.
For example, $G$ a square lattice produces a cubic lattice
statistical mechanical model.

For every Temperley Lieb based statistical mechanical model
which {\em has}  a suitable generalization onto a
lattice with spacelike layer $G$,
such as the Potts model
(defined by Hamiltonian
\begin{equation}
\label{Ham}
{\cal H}   =     
    \sum_{(ij) \in G \times \Z}
        \beta_{ij}\delta_{\sigma_i \sigma_j}
\end{equation}
where $\beta$ is essentially an inverse temperature variable)
 the transfer matrix  algebra provides a
representation (abstractly, a quotient)
of the Full Temperley Lieb algebra.
The inhomogeneous transfer matrix itself is a representation of the element
\begin{equation}
\label{TM}
\label{eqnY}
\tau
=
\prod_{i=1}^n \left(v_i+\sqrt{Q}U_{i.} \right)
\prod_{(ij) \in G} \left(1+\frac{v_{ij}}{\sqrt{Q}}U_{ij} \right)
\end{equation}
where $v=exp(\beta ) -1$.
The Potts representation is given explicitly in \cite{Mar,MarKyo}.
By well known arguments
\cite{Mar,Kogut,SML}
the irreducible representations of $T_G(Q)$
which compose this representation  are then
the
most efficient blocks to use in computing the TM spectrum.
Moreover in two dimensions ($G=A_n$), the exceptional cases,
 where the  algebra
is {\em not} faithfully represented in physical transfer matrices,
 correspond to models with  `rational' conformal field theory
limits. A large amount of information about this conformal limit can actually
be read in the algebraic properties for finite systems \cite{PS,many}.
By establishing the physically correct generic algebra in
other dimensions we develop  a
procedure
for addressing any analogous situation there.

The $G=A_n$ algebra is finite dimensional
for finite $n$, and
typically                          
faithfully represented by the
finite dimensional physical transfer matrices. We will show
that for general $G$ the Full algebra is {\em always} infinite
dimensional unless $G=A_n$.
Since the physical transfer matrices usually
remain finite dimensional in higher dimensions (for finite systems)
one problem is to find
explicitly
the finite dimensional quotients of the Full
algebra appropriate for these physical systems.

We begin (in the next section) with a discussion  of
an algebra related to $T_{\underline{n}}$, called the
partition algebra $P_n(Q)$. It corresponds very closely to the diagram algebra
$D_{\underline{n}}$ which is the
 `mean field limit' of the Potts model.
$P_n(Q)$  also has subalgebras indexed by a graph,
and  is one of the easiest cases to analyse, as expected from a `physical'
point of view.
The algebra $P_n(Q)$ provides an
organisational link between
the physical and abstract algebras we have described.
We begin  analysing $T_G(Q)$ in section 3.  We give a complete analysis of
 the `affine' $\hat{A}_n$ case in section 4.
The  complications that appear here for the Full algebra,
compared to the planar case, may be given a topological interpretation which
leads us in section 5 to a topologically motivated analysis of the general
case. The classification scheme of representations includes
 all links that can be embedded in $G\times\Z$.
Sundry additional remarks are collected  in the last section.

\section{The Partition algebra}

We now discuss
a quotient of the Full algebra
 which will play a crucial role in
our analysis \cite{s2p}.

\subsection{Partitions of a set $M$}

First we need to introduce
the set $S_m$
of
partitions of
 a set $M$ of $m$ distinguished objects
\[
S_{m} = \{
((M_1)(M_2)....(M_i)...)
 :
\hspace{1.5in}
\]
\[
\hspace{1in}
M_i \subseteq M \;\; s.t. \;\; \;\;
M_i \ne \emptyset,
\;\; \;\;
\cup_{i} M_i = M,
\;\; \;\;
M_j \cap M_k = \emptyset \;\; (j \ne k)
\}.
\]
For example, if $M$ is the set of the first $m$ natural numbers
\[
S_2 = \{ ((12)),((1)(2)) \}
\]
\[
S_4 = \{ ((1234)),((1)(2)(3)(4)),
((123)(4)), ((124)(3)),
\hspace{1in}
\]
\[
\hspace{1in}
((134)(2),((234)(1)),
((12)(34)), ((13)(24)), ((14)(23)),
\]
\[
((12)(3)(4)), ((13)(2)(4)), ((14)(2)(3)),
((23)(1)(4)),((24)(1)(3)),((34)(1)(2))
\}.
\]
We call the
individual equivalenced subsets of the set of objects `parts'.
Thus $(M_1)=(123)$ is a part of the partition ((123)(4)), and so on.
The set $S_m$ is finite for finite $m$. The total dimension
is well known
\cite{Mar,MarKyo,Liu}   
\[
\begin{array}{c|cccccccccccccccccc}
m &   1 & 2 &3 & 4&  5&  6&   7&    8    & 9 &    10 &     11 &    12 &    13
\\
\hline
\\
|S_m|&1&2& 5& 15 & 52 & 203& 877 & 4140& 21147 & 115975& 678570& 4213597&
276444
   37
\end{array}
\]
We write $i \sim^A j$
in case objects $i,j $ are in the same partition in $A \in S_m$,
so $\sim^A$ is transitive.

We will be mainly interested in the case $m=2n$.
We will then write our $2n$ objects simply as
\[
1,2,3,...,n,1',2',3',...,n'.
\]

\subsection{Operations on $S_m$  ($m=2n$)}

For      
$Q$ an indeterminate
and $K$ the field of rational functions of $Q$
we define a    
product \cite{Mar,MarKyo}
\begin{equation}
\label{p1}
{\cal P}:
S_m \times S_m \rightarrow KS_m   \hspace{1in}
\end{equation}
\[
(A,B) \mapsto A B = Q^{f(A,B)} C
\]
where $C \in S_m$ and $f(A,B) \in \Z_{\ge 0}$ are defined
as follows.
Relabel the objects in $B$ from
\[
1'',2'',..,n'',1''',2''',..,n'''.
\]
Form a partition of $4n$ objects from $A$ and $B$ by first taking
the parts in $A$ and including into the part
containing  $i'$ the part from $B$ containing $i''$.
 Then delete all the $i'$ and $i''$ (counting the number
$f(A,B)$ of parts which become empty,
and are then discarded,
 in the process)
and finally relabel all the $i'''$ as $i'$. The
partition of $2n$ objects obtained is $C$.

For example,
\[
((1234)(1'3')(2')(4')) \;\; ((11'2')(233')(44'))
\rightarrow
((1234)(1'3'1''1'''2'''2''3''3'''2')(4'4''4'''))
\]
\[
\hspace{1in}
\rightarrow
((1234)(1'''2'''3''')(4'''))
\rightarrow
((1234)(1'2'3')(4')).
\]
This is illustrated in figure~\ref{di1}. There are
other such illustrations in section~\ref{spic}.

\begin{figure}
\label{di1} %
\begin{picture}(300,220)(-140,-220)
\put(-60,-20){$\huge A$}
\put(-60,-100){$\huge B$}
\put(0,2){1}
\put(40,2){2}
\put(80,2){3}
\put(120,2){4}
\multiput(0,0)(0,-80){2}{
\put(60,-20){\oval(180,70)}
}
\multiput(0,-80)(40,0){4}{\line(0,1){40}}
\thicklines
\multiput(0,0)(0,-40){4}{                    
\multiput(0,0)(40,0){4}{\circle*{5}}
}
\put(0,0){\line(1,0){120}}
\put(40,-40){\oval(80,15)[t]}
\put(40,-80){\line(1,0){40}}
\put(0,-120){\line(1,0){40}}
\multiput(80,-80)(40,0){2}{\line(0,-1){40}}
\multiput(0,-80)(10,0){1}{\line(0,-1){40}}
\put(60,-150){$\downarrow$}
\multiput(0,-170)(0,-40){2}{                    
\multiput(0,0)(40,0){4}{\circle*{5}}
}
\put(0,-170){\line(1,0){120}}
\put(0,-210){\line(1,0){80}}
\thinlines
\put(60,-190){\oval(180,70)}
\put(-60,-190){$\huge C$}
\end{picture}
\caption{Pictorial realisation of parts as clusters
and composition of partitions by juxtaposing clusters. In this case
$f((A,B))=0$.}
\end{figure}
\begin{figure}
\label{d3graph} 
\begin{picture}(300,190)(-140,-40)
\multiput(0,0)(0,40){4}{                    
\multiput(0,0)(40,0){2}{\circle*{5}}
\multiput(20,20)(40,0){1}{\circle*{5}}
\put(0,0){\line(1,0){40}}
\put(0,0){\line(1,1){20}}
\put(20,20){\line(1,-1){20}}
\multiput(0,0)(40,0){2}{\line(0,-1){40}}
\multiput(20,20)(10,0){1}{\line(0,-1){40}}
                                            }
\end{picture}
\caption{Part of the graph $\hat{A}_3 \times \Z =
\underline{3} \times \Z$.}
\end{figure}
\begin{figure}
\label{d3perm}
\begin{picture}(300,190)(-140,-40)
\multiput(0,0)(0,40){4}{                    
\multiput(0,0)(40,0){2}{\circle*{5}}
\multiput(20,20)(40,0){1}{\circle*{5}}
\put(0,0){\line(1,0){40}}
\put(0,0){\line(1,1){20}}
\put(20,20){\line(1,-1){20}}
\multiput(0,0)(40,0){2}{\line(0,-1){40}}
\multiput(20,20)(10,0){1}{\line(0,-1){40}}
                                            }
\thicklines
 \multiput(1,120)(40,0){2}{\line(0,-1){40}}
\put(1,82){\line(1,1){20}}
\put(21,101){\line(0,-1){80}}
\put(41,81){\line(0,-1){40}}
\put(1,41){\line(1,0){40}}
\put(21,20){\line(1,-1){20}}
   \multiput(1,1)(40,0){2}{\line(0,-1){40}}
\put(40,130){$2$}
\put(20,150){$3$}
\put(0,130){$1$}
\put(45,-35){$2'$}
\put(25,-15){$3'$}
\put(5,-35){$1'$}
\put(1,1){\line(0,1){40}}
\end{picture}
\caption{Diagram for the connectivity
               $1_{12}U_{3.}=((12')(21')(3)(3'))$
which restricts to $1_{12}$ for $n=2$.}
\end{figure}

\begin{de}[Partition algebra \cite{s2p}]
Considering the vector space over $K$   
spanned by $S_{2n}$,
the linear extension of
the product ${\cal P}$ gives us a
finite dimensional
 algebra over $K$     
which we
call the partition algebra $P_n(Q)$.
\end{de}

\begin{de}
The {\em natural} inclusion $\cal S$ is defined by
\[
0 \rightarrow P_{n-1} \stackrel{\cal S}{\rightarrow} P_n
\]
\begin{equation}
\label{eqn15}
{\cal S}:((...)...(..)) \mapsto ((...)...(..)(nn'))
{}.
\end{equation}
\end{de}

It is convenient to introduce the following special elements of the
partition algebra:
\begin{equation}
\label{ee1}
1=((11')(22')...(nn'))
\end{equation}
\begin{equation}
\label{ee2}
1_{ij} = ((11')(22')..(ij')..(ji')..(nn'))
\hspace{.5in} i,j =1,2,..,n
\end{equation}
\begin{equation}
\label{ee3}
A_{i.} = \frac{1}{\sqrt{Q}} \;\; ((11')(22')...(i)(i')...(nn'))
\end{equation}
\begin{equation}
\label{ee4}
A_{ij}= \sqrt{Q} \;\;  ((11')(22')...(iji'j')...(nn')).
\end{equation}
\begin{pr}[see  \cite{s2p}]
\label{pro6}
These elements generate $P_n(Q)$.
\end{pr}
\begin{de}
For $A \in P_n$
let $[A]$ denote the
maximum over the $S_m$ components of $A$ of the
number of distinct parts containing both
primed and unprimed elements.
\end{de}
For example $[1]=n$, $[A_{i.}]=n-1$.
Then
\begin{co}
For $A,B \in P_n$
\[
[AB] \le  min([A],[B])
{}.
\]
\end{co}

\begin{pr}
\label{pro7}
There is a homomorphism from the Full Temperley-Lieb algebra to
the partition algebra given by
\[
H: T_{\underline{n}}(Q) \rightarrow P_n(Q)
\]
\[
H:1 \mapsto 1
\]
\[
H:U_{i.} \mapsto A_{i.}
\]
\[
H:U_{ij} \mapsto A_{ij}.
\]
\end{pr}

%
\begin{pr}[see \cite{Mar}]
\label{fred}
The subalgebra of $P_n(Q)$ generated by
\[
<1, \;\; A_{i.} \;\; (i=1,2,..,n)
, \;\; A_{ii+1}  \;\; (i=1,2,..,n-1)>
\]
is isomorphic to $T_{A_n}(Q)$.
\end{pr}

\begin{de}
For given $n$ we define
$\Sigma_n$  as the subalgebra of $P_n(Q)$ generated
by
\[
< 1, 1_{ij} \;\; (i,j=1,2,...,n)>
\]
or, where appropriate, as the corresponding symmetric group.
\end{de}

\subsection{Full embedding of $P_{n-1}$ in $P_n$}
\label{sss2.1}
\label{sfet}
We will need the following simple but surprisingly powerful theorem:
\begin{th}[see \cite{s2p}]
\label{fet}
For each $n$, $Q\ne 0$ and idempotent $e=e_n=A_{n.}/\sqrt{Q}$
 there is an isomorphism
of algebras
\[
e_n P_n e_n \cong P_{n-1}
{}.
\]
\end{th}
%

As a consequence
the categories of left $P_{n-1}$ and left $P_ne_nP_n$ modules are
essentially isomorphic categories (the extent to which they are
not isomorphic is, for our purposes, a technicality - the
interested reader should turn to
\cite{s2p,Cline,bww} for
details.

Let us denote by
$F_n(M)=e_nM$   
the object map from $(P_n-mod)$ to $(P_{n-1}-mod)$
at level $n$.

\begin{cor}[see \cite{s2p}]
Let $f_n$ be the object map of categories defined by
restriction of left $P_n$ modules to  left 
$P_{n-1}$ modules through the inclusion $\cal S$,
\[
f_n: ( P_n -mod ) \rightarrow   ( P_{n-1} - mod)
\]
\[
f_n : M \mapsto {}_{{}_{P_{n-1}}} \! \downarrow  \! M
{}.
\]
 Then
the following diagram of object maps of categories commutes:
\begin{equation}
\begin{array}{cccccc}
(P_n  -mod ) & \stackrel{F_n}{\longrightarrow} & 
                                                 ( P_{n-1} -mod )
\\
\\
{}_{ f_n} \downarrow &                    & \downarrow_{ f_{n-1}}
\\
\\
      (P_{n-1} -mod)     & \stackrel{F_{n-1}}{\longrightarrow} & (P_{n-2} -mod)
\end{array}
{}.
\end{equation}
\end{cor}

This implies that, up to edge effects caused by the difference between
$P_n$ and $P_ne_nP_n$, the Bratteli restriction diagram for
the algebras $P_n$
(see section~\ref{sX} onwards)
has the same structure on each level $n$. But then

\begin{pr}[see \cite{s2p}]
The following is a
short exact sequence of algebras
\[
0 \rightarrow P_neP_n \rightarrow P_n \rightarrow
\Sigma_n
\rightarrow 0
{}.
\]
\end{pr}

Thus, at least for $P_n$ semi-simple, a knowledge of
the structure of $P_{n-1}$
essentially determines for us the structure of $P_n$.
\begin{co}
In case $P_n(Q)$ semi-simple 
the distinct equivalence
classes of irreducible representations
may be indexed by the list of all standard partitions of every integer from $0$
(understood to have one standard partition) to $n$.
\end{co}

In fact  $P_n(Q)$ is
semi-simple for $Q$ indeterminate and for all $Q \in \C$ except for
the roots of a finite order polynomial in $Q$ for any finite $n$. See section
6.


\subsection{Diagram algebra for a graph $G$}
Let us return to proposition~\ref{fred}.
More generally we have
\begin{de}
For graph $G$ the Diagram algebra $D_G(Q)$
is defined as the subalgebra of the partition algebra
generated by
\[
<1,\;\; A_{i.} \;\; (i=1,2,..,n)
, \;\; A_{ij}  \;\; (i,j \in G)>.
\]
\end{de}
Note that $D_{\underline{n}}(Q) \subset P_n(Q)$, as $1_{ij}$
cannot be built with these generators.
However, under certain conditions it can be substituted, for example,
\begin{equation}
\label{eex}
1_{23} A_{1.} = A_{1.} A_{12} A_{2.} A_{23} A_{3.} A_{13} A_{1.}
{}.
\end{equation}
In fact
we are more interested here in
 $D_{\underline{n}}(Q) $ than $ P_n(Q)$
(compare proposition~\ref{pro7} with equation~\ref{eqnY}),
but $P_n(Q)$ provides a more versatile   
general setting.
We will see shortly that it is straightforward to move from
one to the other.

The relationship between the algebra types
$T,P$ and $D$ is summarized by  the commutative  diagram
\[
\begin{array}{ccccccccccccccc}
  &                                  &0
\\
  &                                  & \downarrow
\\
T &  \stackrel{H_1}{\longrightarrow} &D          & \longrightarrow & 0
\\
  &{}_H \searrow                     &\downarrow_{H_2}
\\
  &                                  &P
\end{array}
\]
which is  exact at $D$.

\begin{pr}
\label{pr3}
The subalgebra $D_{\underline{n}}(Q) \subset P_n(Q)$ is invariant under
conjugation by elements of the group $\Sigma_n$, i.e.
\[
b^{-1} D_{\underline{n}}(Q)   b =  D_{\underline{n}}(Q)
\;\;\; \forall b \in \Sigma_n.
\]
\end{pr}

\begin{co}
Every word in $P_n(Q)$ can be written in the form $AB$ where
$A \in \Sigma_n$ and $B\in D_{\underline{n}}(Q)$.
\end{co}

Clearly we have the rich inclusion structure again
\[
G \supset G' \;\; \Rightarrow \;\;  D_G(Q) \supseteq D_{G'}(Q).
\]

It also follows that $D_G(Q)$,
and indeed $P_n(Q)$, obeys a number of quotient
relations in addition to the Temperley-Lieb relations. For
example, with $W \in D_G(Q)$
there exists $X(W)$ a certain (known)
scalar function of $Q$ (see \cite{Mar})
such that
\[
\left( \prod_i A_{i.} \right) \;\;
W  \;\;
\left( \prod_i A_{i.} \right)
=
X(W) \;\;
\left( \prod_i A_{i.} \right).
\]
Specifically, if $W \in S_m$ with
$b_W$ parts
\[
X(W) = Q^{b_W}.
\]
This relation is
suitable for
at least {\em part} of the set
appropriate for physical systems, as it corresponds to
the existence of disorder at very high temperatures
(there is also a dual corresponding to order at low temperatures).
At the level of the dichromatic polynomial it corresponds to
isolating $b_W$ clusters (c.f. \cite{Baxter}, for example).
Several analogous relations have also been found \cite{Mar}.

\subsection{Pictorial realisation by Connectivities}
\label{spic}

\begin{de}
For a graph $G$ let ${\cal B}_G$
be the universal set of the set of bonds of $G$,
i.e. the set of all (not necessarily proper)
subgraphs of $G$ of order $|G|$ nodes (obtained by omitting bonds).
\end{de}
Note that elements of ${\cal B}_G$ may not be connected graphs
\cite{Baxter,blote}.

The partitions $A \in D_{\underline{n}}(Q) \cap S_m$
 may be realised as  classes of
 ${\cal B}_{\underline{n} \times \Z}$
   either
under a certain equivalence $\rho$. The construction of $\rho$ is as
follows.

Explicitly number the nodes of $\underline{n}$ at `time' $t=0$
from $1,2,...,n$.
Practically it will be convenient
to restrict attention in
${\cal B}_{\underline{n} \times \Z}$
to the subset of elements which for
sufficiently large $t$ have all time-like bonds present
and all space-like bonds absent. This is a sort of very large time
asymptotic smoothness condition.
For some such very large $t=T$ number the nodes
of $(\underline{n},T)$ from $1',2',...,n'$.
Then introduce the map
\[
F_T: {\cal B}_{\underline{n} \times \Z} \rightarrow P_n(Q)
\]
\[
F_T: B_o \mapsto Q^b \; B
\]
where $B \in S_m$
such that $i \sim^B j$ iff $i,j$ are connected by a path of bonds
present in the subgraph $B_o$,
and $b$ is the number of isolated connected components in $B_o$
not connected to any point in either of the layers
$t=0$ or $t=T$.

The point about the limits $t=0,T$ is that
for finite $n$ there exists some finite $T$
beyond which  (range$F_T$)$ \cap S_m$ does not increase.
Thus the asymptotic condition is not important (just convenient),
but ensures that $F_T$ and $F_{T+1}$ are essentially the same map.

The
equivalence classes of ${\cal B}_{\underline{n} \times \Z}$
are defined so as to make this map an injection
(i.e. $B_o \rho C_o $ only if $B=C$).

The range of $F_T$ does not include the whole of $S_m$
however large we make $T$
(see the remark after definition~1).
We can
extend to the whole of $S_m$ by, for example, building our
`connectivities' on $\underline{n+1} \times \Z$ (but only
labelling the `first' $n$ nodes).
This complication is connected to the nature of the
lattice and the TM formalism, it will be discussed further elsewhere.
In general,
different choices
of $G$ in
${\cal B}_{G \times \Z}$,
realise
different sets of conectivities, i.e. different ranges for $F_T$.
This is, in fact, the essence of the
physically important problem of finding irreducible
representations of $D_G(Q)$ (see later).

We may extend ${\cal B}_{\underline{n+1} \times \Z}/\rho$ or
 ${\cal B}_{\underline{n} \times \Z}/\rho$ to an algebra
(over rational functions in $Q$) by defining a product
$B_o \; C_o$.
We first build a new graph $(BC)_o$ by
discarding $t>T$ in $B_o$ and $t<0$ in $C_o$ and then
joining $B_o$ and $C_o$
 by identifying the layer $t=T$ in $B_o$ with $t=0$ in $C_o$.
Let $D \in S_m$ have the same connectivities
as the graph so produced has between $t=0$ and $t=2T-1$
(i.e. $F_{2T-1}((BC)_o)=Q^d D$ for some $d$).
We then define
$B_o C_o = Q^d D_o$ where $D_o \in {\cal B}/\rho$
is such that $F(D_o)=D$.
The map $F$ is then an algebra homomorphism.

The explicit pictorial realization is particularly neat (but
sufficiently general) if we
distribute the nodes of $\underline{n}$
linearly, as in $A_n$, and
only draw the part of the graph not in the asymptotic region.
Then for example
with $n=12$ the $\rho$ class of $A_{ii+1}$ has a simple
representative

\begin{picture}(300,30)(-120,-10)
\put(-75,0){$A_{i \; i+1} /\sqrt{Q}  \;\; \leftarrow$}
\multiput(0,0)(10,0){12}{\circle*{3}}
\put(70,0){\line(1,0){10}}
\put(70,5){$i$}
\end{picture}

\noindent
The $\rho$ class of $A_{i.}$ has representative

\begin{picture}(300,40)(-120,-20)
\put(-70,0){$\sqrt{Q} \; A_{i.}  \;\; \leftarrow$}
\multiput(0,0)(10,0){12}{\circle*{3}}
\multiput(0,0)(10,0){7}{\line(0,-1){10}}
\multiput(80,0)(10,0){4}{\line(0,-1){10}}
\multiput(0,-10)(10,0){12}{\circle*{3}}
\put(70,5){$i$}
\end{picture}


The composition rule is to identify the top row of dots in the second
diagram with the bottom row in the first. Clusters then isolated from
both top and bottom rows of
the new diagram so formed may be removed, contributing a factor $Q$.

Finally, then, for example, the TL relation~\ref{r5}
\[
A_{i \; i+1} A_{ i.}  A_{i \; i+1}= A_{i \; i+1}
\]
amounts to  the statement that
the subgraph

\begin{picture}(300,40)(-100,-20)
\multiput(0,0)(10,0){12}{\circle*{3}}
\multiput(0,-10)(10,0){12}{\circle*{3}}
\put(70,0){\line(1,0){10}}
\put(70,-10){\line(1,0){10}}
\multiput(0,0)(10,0){12}{\circle*{3}}
\multiput(0,0)(10,0){7}{\line(0,-1){10}}
\multiput(80,0)(10,0){4}{\line(0,-1){10}}
\multiput(0,-10)(10,0){12}{\circle*{3}}
\put(70,5){$i$}
\end{picture}

\noindent
has the same list of connections within and
between the top and bottom layers as the
$\rho$ representative of $A_{i\; i+1}$ above.

Note that no 
composition of diagrams
increases the number of {\em distinct}
connected clusters
connecting between the top and bottom layers.
This means that the subset of $\rho$ cosets with no
connections top to bottom  
form a basis for a
$P_n(Q)$ bimodule.
Furthermore, the subset with $\le p$ distinct
connections top to bottom also form a basis for
a $P_n(Q)$ bimodule.

\subsection{Structure and Representation Theory of $P_n(Q)$}
\label{s2.3}

This picture is particularly useful for constructing representations.
The number of distinct connections running from $t=0$ to $t=T$
is evidently
non-increasing in any composition
(it is a measure of the number of distinct
bits of information which can be simultaneously
propagated through the bond covering,
which cannot exceed  the number propagated
across any fixed time slice).
So for example,
writing simply $P$ or $P_n$ for $P_n(Q)$, and
defining idempotents
\[
I_k = \prod_{i > k} \frac{H(U_{i.})}{\sqrt{Q}}
\]
($Q \ne 0$) then
$I_0$ allows {\em no} connections from $t=0$ to $t=T$, so
$P_n I_0 P_n$ is the invariant subspace of $P_n$ where
\[
\not \! \exists \;\;  A,i,j \;\; s.t. \;\; i \sim^A j'.
\]

\begin{pr}
The element $I_0$ is a primitive idempotent.        
   n).
\end{pr}
\begin{co}
The left ideal $P_n I_0$ is indecomposable (and generically simple).
\end{co}
Note that $dim(P_n I_0)=|S_n|$.
\begin{pr}
The element $I_1$ is primitive in the quotient algebra $P_n/P_nI_0P_n$.
\end{pr}
so again $P_nI_1$ is indecomposable in this quotient.

Now $I_2$ is not primitive in $P_n/P_nI_1P_n$ since, for example
\[
I_2 1_{12} I_2 \sim 1_{12} I_2 \not\propto I_2
{}.
\]
On the other hand $\frac{(1+1_{12})}{2}I_2$ and $\frac{(1-1_{12})}{2}I_2$ are
primitive idempotents.      

Similarly $I_3$ is not primitive in $P_n/P_nI_2P_n$, but, for example
\[
\Sigma_{\pm} I_3 =
\frac{(1 \pm 1_{12} \pm 1_{23} \pm 1_{13}+1_{12}1_{23}+1_{13}1_{23})}{3!}I_3
\]
and two further combinations (with other symmetries) are.

{}From the definition of $I_i$ we have
$P_nI_{i-1}P_n \subset P_nI_iP_n$ and
a nest of short exact sequences, $i=1,2,...,n$  
\[
0 \rightarrow P_nI_{i-1}P_n \rightarrow P_nI_{i}P_n \rightarrow
P_nI_iP_n/P_nI_{i-1}P_n \rightarrow 0
\]
where finally $I_n=1$.
\begin{de}
Let us define the algebra $P_n[i]=P_nI_iP_n/P_nI_{i-1}P_n$.
\end{de}
This is the algebra
of elements with not more than $i$ distinct connections
running, as it were, from $t=0$ to $t=T$, quotiented by
the invariant subspace of all elements with strictly
less than $i$ distinct connections from $0$ to $T$.
\begin{pr}
In the quotient $P_n[i]$
\[
I_i \Sigma_n I_i = \Sigma_i \; I_i
\]
(we take $\Sigma_0=\Sigma_1=1$).
\end{pr}

\begin{pr}
\label{pr7}
Let $\Sigma$ be any left $\Sigma_i$ module. Then we
 can write the left $P_n[i]$  
 module
\[
P_n(Q) \left( I_i \Sigma \right)
 =
 D_{\underline{n}}(Q) \left( I_i \Sigma \right)
\]
\end{pr}

\subsubsection{
To construct irreducible representations:}

For each $i=0,\ldots,n$ and $\lambda\vdash i$ ($\lambda$ a
partition of $i$) and $\Sigma_{\lambda}$ the
$\lambda$ simple $\Sigma_i$ module \cite{GdeB}
, the set $S_mI_i\Sigma_{\lambda}$ generates a basis
for a generic irreducible representation.

Let us first consider
the
fully symmetrized case
for the left $\Sigma_i$ module,
call it $\Sigma^s$, in each
sector $i$.
Then we get a basis for
the left $P_n[i]$ module $P_n  I_i \Sigma^s$  
from $S_m$ as follows.
List the elements as partitions of $1,2,...,n$, ignoring $1',2',...,n'$
except in so far as to note which  parts
originally contained primed elements.
We discard duplicate copies of
partitions not distinguished by this property,
and partitions in which other than $i$ parts originally
contained primed elements.
We call the resultant set $S_n(i)$.
For example,
\[
S_2(1)= \{ ((12)'), ((1)'(2)), ((1)(2)') \}
{}.
\]
We do not need to keep track of exactly {\em which}
unprimed nodes 
 were connected to
{\em which} primed nodes,     
since the symmetriser makes all these permutations equivalent.
In other words  the set $S_n(i)$
is the set
of all possible ways of
arranging the elements of $S_n$ (c.f. $S_m=S_{2n}$)
so that $i$ parts are
distinguished from the rest.
An element of $S_n$ with $p \ge i$ parts produces
$p!/((p-i)!i!)$ elements of the basis $S_n(i)$
(and produces none if $p<i$).    
Note that
\[
\sum_{i=0}^n S_n(i) = 2^n S_n.
\]
The action of the generators on such a basis is
just the usual product (\ref{p1}) (remembering the $P_n[i]$ quotient).
It is given explicitly  in \cite{Mar,MarKyo}.
We will prove irreducibility of these representations in section~\ref{s2.3.2}.

Moving to the case where we take some other left $\Sigma_i$ module
in proposition~\ref{pr7}, then
our basis must simply be
(semi) direct producted with a basis for this new module.
Permuting actions will act on the $\Sigma_i$ module rather than the partitions.

\subsubsection{The case $n=3$}

We can well illustrate all of these points with an example.
Let us consider $n=3$. The available partition shapes $\lambda$
in $S_6$ are:
\[
(6),(5,1),(4,2),(3^2),(4,1^2),(3,2,1),(2^3),(3,1^3),(2^2,1^2),(2,1^4),
(1^6)
\]
with corresponding multiplicities ${\cal D}_{\lambda}$:
\[
1,6,15,10,15,60,15,20,45,15,1
\]
giving total dimension $|S_6| = 203$.

On the other hand the dimensions of the bases described above are
\[
5,10,6 \; dim(\Sigma_2),1 \; dim(\Sigma_3)
\]
i.e., explicitly, the bases are (with parts after the colon
understood primed)
\[
\{((123):\emptyset)
,((12)(3):\emptyset)
,((13)(2):\emptyset)
,((23)(1):\emptyset)
,((1)(2)(3) :\emptyset)
\},
\]
\[
\{
(\emptyset
:(123))
,
((12):(3))
,((3):(12))
,((13):(2))
,((2):(13))
,
\hspace{1in}
\]
\[
\hspace{1in}
((23):(1))
,((1):(23))
,
((1)(2):(3))
,((1)(3):(2))
,((2)(3):(1))
\}
,
\]
\[
\{ (\emptyset
:(12)(3)),
(\emptyset:(23)(1)),
(\emptyset:(2)(13)),
((1):(2)(3)),
((2):(1)(3)),
((3):(1)(2))  \} \times \Sigma_{\pm}
\]
\[
\{(\emptyset:(1)(2)(3)) \} \times \Sigma_3
\]
Finally, then, noting the multiplicities of    
 inequivalent generically irreducible representations
at level $i$ we have
\[
5^2 + 10^2 +6^2 .(1+1) +1^2 . (1+2^2+1) =203=\left|S_6\right|
\]
 so we have, for example,
 the complete set of
inequivalent irreducible representations for the semi-simple cases.
Note that all the $i=3$ representations reduce to (direct sums of) the
same representation in $D_{\underline{n}}(Q)$, because none of the
permutations can actually be realized in this subalgebra.

\subsubsection{General $n$}
\label{s2.3.2}
\label{sX}
\label{sss2.4.3}

Since we know the structure of
the symmetric group (algebra) $\Sigma_i$
(see, for example, \cite{Hamermesh,GdeB})
it behoves us to
divide up our analysis by first considering
 the algebra for the $\Sigma$-symmetrised case, $P_n(Q)/\sim$,
which we define below.
The rest then follows from
changing the left $\Sigma_i$ module in propostion~\ref{pr7}.
\begin{de}
We define an equivalence relation $\sim$ on $S_m$ by $A \sim B$ iff
they
are the same up to a permutation of the
connections made by the connectivities from $t=0$ to $t=T$.
\end{de}

We write $P_n(Q)/\sim$ for the quotient algebra
obtained by the linear extension
to $P_n(Q)$.

In this case there is a bra-ket notation for elements of $S_m$. Every
element may be written uniquely in the form $|a><b|$ where $a,b \in S_n(i)$
for some $i$ (conversely every
such pair defines a unique element).
 There is then an obvious inner product $<b|c>$
in each $P_n[i]/\sim$, obtained
from
\[
|a><b||c><d| = <b|c> \; |a><d|
{}.
\]
\begin{pr}
The $n+1$ representations
of $P_n(Q)/\sim$ with bases $S_n(i)$ ($i=0,1,2,..,n$)
and canonical action (up to the $P_n I_{i-1} P_n$ quotient)
are each irreducible for $Q$ indeterminate.
\end{pr}
\begin{co}
These representations are inequivalent.
\end{co}
\begin{co}
\label{co10.2}
Any representation of $P_n(Q)$ built from proposition~\ref{pr7}
with $\Sigma$ an irreducible $\Sigma_i$ module is irreducible.
\end{co}

\begin{co}
$P_n(Q)$ is semi-simple for $Q$ indeterminate and for all $Q \in \C$
except for the roots of a finite order polynomial in $Q$ for any finite $n$.
\end{co}
{\em Proof:}
by dimension counting. The irreducible
representations account for the full
dimension of the algebra. We can show this explicitly as follows:

The Bratelli diagram for the restriction corresponding to
\[
\left( P_n(Q)/\sim \right) \supset \left( P_{n-1}(Q)/\sim \right)
\]
on these    
irreducible representations
is as follows (with top line $n=0$)
\newcommand{\nw}{\nwarrow}
\newcommand{\nea}{\nearrow}
\newcommand{\nx}{\nw \!\!\!\!\!\! \nea}
\newcommand{\nee}{\nx \!\!\!\! \nea}
\newcommand{\neee}{\nee \!\!\!\! \nea}
\newcommand{\neeee}{\neee \!\!\!\! \nea}
\newcommand{\neeeee}{\neeee \!\!\!\! \nea}
\newcommand{\up}{\uparrow}
\newcommand{\upp}{\up \! \up}
\newcommand{\uppp}{\upp \! \up}
\newcommand{\upppp}{\uppp \! \up}
\newcommand{\uppppp}{\upppp \! \up}
\newcommand{\upsix}{\uppppp \! \up}
\[
{\large
\begin{array}{cccccccccccccccccccccc}
1
\\
\up & \nw
\\
1   &     & 1
\\
\up & \nx & \upp & \nw
\\
2   &     & 3   &       &    1
\\
\up & \nx & \upp& \nee &    \uppp & \nw
\\
5   &     &  10 &       &    6   &       &    1
\\
\up & \nx & \upp& \nee  & \uppp  &  \neee& \upppp & \nw
\\
15  &     &  37 &       &  31    &       &   10   &        &  1
\\
\up & \nx & \upp& \nee  & \uppp  &  \neee& \upppp & \neeee &\uppppp&\nw
\\
52  &     &  151&       &  160   &       &   75   &        &  15  &&1
\\
\up & \nx & \upp& \nee  & \uppp  &  \neee& \upppp & \neeee
&\uppppp&\neeeee&\ups
   ix&\nw
\\
203 &     &  674&       &  856   &       &  520   &        & 155  &&21&&1
\end{array}
}
\]
and so on. These restrictions are forced by a Morita equivalence
theorem - c.f. \cite{Mar}.
This works as follows. The isomorphism of
categories in the corollary to theorem~\ref{fet}
takes a layer of the above diagram to the layer below it (each node is mapped
vertically down, since the idempotent $e_n$ cuts at most one connection).
The $1$ at the right
hand side of the lower layer is missing, of course, as this is the
trivial representation of $\Sigma_n$. Consequently
(i.e. as a knock on effect from the {\em previous} layer) the
restriction information for the next two modules to the left is
incomplete.
However, the only possibility is for the
restrictions to include some copies of the
trivial representation, and  these may be filled in by dimension counting.
 For example, omitting node 3 in
$S_3(2)$ we get
\[
\{
((1)'(23)'),((12)'(3)'),((13)'(2)'),((1)'(2)'(3)),((1)'(2)(3)'),((1)(2)'(3)')
    \}
\]
\[
\rightarrow
\{
((1)'(2)'),((12)'),((1)'(2)'),((1)'(2)'),((1)'(2)),((1)(2)') \}
=S_2(1) + 3.S_2(2)
{}.
\]
Note that in omitting the last node ($n$) in this mnemonic
if we have a part of the form $(ij...mn)$
(i.e. unprimed) then this maps to $(ij...m)'$,
since the action of generators here is as if the part
is connected to {\em something}!

Irreducible representations of $P_n(Q)$ follow by corollary~\ref{co10.2}.

Let us write $d_n(i)$ for the
dimension of the $i^{th}$
representation in row $n$ (the $i^{th}$ column, counting the left hand
column as column 0). Then the total dimension of $P_n(Q)$ is
\[
|S_m|
=
\sum_{i=0}^n (i)! (d_n(i))^2   =  |S_m(0)|
\]
when $m=2n$
as required
(the last identity is readily proved). Altogether $S_m(i) \times
\Sigma_{\lambda}$ ($\lambda \vdash i$) gives
the complete structure for all semi-simple cases.

\section{$D_{G}$ for arbitrary graphs $G$}

\subsection{The structure of $D_{\underline{n}}(Q)$}
The structure of the $D_{\underline{n}}$ algebra is very similar to $P_n(Q)$.
For $P_n$ each node
in column $i$ of the Bratelli
diagram above represents a list of irreducibles, one for
each partition of $i$ (with dimension muliplied by the
dimension of the corresponding representation of the
symmetric group). The only difference here is that the $i=n$ representation
(the rightmost one on each line) gives only a single
(one dimensional) representation. This just
corresponds to the impossibilty of any transverse
movement of $n$ distinct connected lines on a
graph with only $n$ nodes in the lateral direction.

\subsection{On the structure of $D_G$}

\begin{pr}
The connectivities which can be realised on $G\times\Z$ provide bases for the
irreducible representations of $D_G$.
\end{pr}

This follows since in $D_G$ these subsets span  invariant subspaces. Computing
the dimension is more difficult. See section 6.

\vspace{.2in}

We will find in section 4 that the  above
 finite dimensional
 algebras do not include all those appropriate for  systems of interest in
physics.

Two important questions arise:

1. How do the finite dimensional algebras $D_G$
 fit into the overall structure of $T_G$? - see section 4.

2. What is the asymptotic growth rate of dimensions of
representations corresponding to a given physical observable
 or more generally,
 which $Q$ values are exceptional for each physical
dimension? This is also answered in some cases in section 4.

\subsection{Review of $G=A_n$}

Consider a twice punctured sphere. Put the unprimed elements of $M$
in 
natural order at intervals around one boundary, and the
primed ones
opposite them
 around the other boundary.
Draw a `seam' from one boundary to the other on the sphere, begining
between 1 and $n$ and ending between $1'$ and $n'$.
A partition $A \in S_m$ is accessed from $T_{A_n}(Q)$
(i.e. in $H(T_{A_n}(Q))$)
iff for every part in $A$ a line can be drawn on the sphere connecting
all the elements of the part, but touching no other lines,
and not touching the seam.

If the connecting lines are thickened into ribbons then
the usual boundary diagrams \cite{Mar,bww,Kauf}
 - figure~\ref{Uipic}
and so on, in case $n=6$ - represent the
edges of these lines.
Representatives of all
 equivalence classes under continuous
deformation of non-overlapping segments
 are realized.

A boundary
diagrammatic representation of words in
$D_{A_n}(Q)$ is obtained as follows.
The generator $U_i$  maps to the coset of
the diagram shown in
figure~\ref{Uipic}.
\begin{figure}
\begin{picture}(100,100)(-100,30)
\multiput(0,100)(10,0){5}{\line(0,-1){60}}
\put(55,100){\oval(10,20)[b]}
\put(55,40){\oval(10,20)[t]}
\put(-2,103){1}
\put(8,103){2}
\put(18,103){..}
\put(47,103){i}
\put(54,103){i+1}
\put(67,103){ }
\put(87,103){..}
\put(106,103){n}
\multiput(70,100)(10,0){5}{\line(0,-1){60}}
\end{picture}
\caption{Boundary
diagram for $U_i$ in $D_{A_n}(Q)$, c.f. $A_{j.}$ ($j=(i+1)/2$)
in the connectivity picture. \label{Uipic}}
\end{figure}
Generators
 are then composed by connecting the bottom of the diagram
for the first factor with the top of the diagram for the second,
so
\[
U_i U_{i+1} U_i = U_i
\]
 is given by figure~\ref{uuu}.
\begin{figure}
\begin{picture}(100,140)(-100,-10)
\multiput(0,100)(10,0){5}{\line(0,-1){100}}
\put(55,100){\oval(10,20)[b]}
\put(55,70){\oval(10,20)[t]}
\put(65,70){\oval(10,20)[b]}
\put(65,35){\oval(10,20)[t]}
\put(55,35){\oval(10,20)[b]}
\put(48,103){i}
\put(55,00){\oval(10,20)[t]}
\put(70,100){\line(0,-1){30}}
\put(50,70){\line(0,-1){35}}
\put(70,35){\line(0,-1){35}}
\multiput(80,100)(10,0){4}{\line(0,-1){100}}
\end{picture}
\caption{Boundary diagram exhibiting
the relation $U_iU_{i+1}U_i=U_i$. \label{uuu}}
\end{figure}
In such diagrams we simply pull the line starting in the $i+2$
position straight to exhibit the identity.

Similarly
$U_i U_i = \sqrt{Q} U_i$ is represented in figure~\ref{sq}.
\begin{figure}
\begin{picture}(100,110)(-100,10)
\multiput(0,100)(10,0){5}{\line(0,-1){80}}
\put(55,100){\oval(10,20)[b]}
\put(55,20){\oval(10,20)[t]}
\put(55,60){\oval(10,20)}
\multiput(70,100)(10,0){5}{\line(0,-1){80}}
\end{picture}
\caption{Diagram for $U_iU_i$
before removing the closed loop. \label{sq}}
\end{figure}
Here we must interpret closed loops as removable, after contributing a factor
$\sqrt{Q}$.  These diagrams can then be
used to match any word to
a reduced word (one which cannot be shortened using the
relations), with an appropriate scalar factor.
\label{s5.2}

\vspace{1in}

Note that here
\[
I_0 = \prod_{i} U_{i}/\sqrt{Q}
\]
($Q\ne 0$)
is a primitive idempotent.                 
Writing simply $T$ for $T_{A_n}(Q)$
the left ideal $TI_0$
is thus indecomposable (and hence generically
irreducible).
Let us define
\[
I_k = \prod_{i > k} \frac{U_i}{\sqrt{Q}}
{}.
\]
 On quotienting by
the double sided ideal $TI_0T$
we find that  $I_1$
becomes a primitive idempotent,
so $TI_1$ is indecomposable in $T/TI_0T$, and so on.
Iterating this filtering process with respect to $k$
we generate bases for all $n+1$ generically irreducible
representations.
The dimensions of these representations are then readily determined
by reference to the tower relation
\begin{equation}
\label{tower}
A_n \subset A_{n+1} \;\; \Rightarrow \;\; T_{A_n}(Q) \subset T_{A_{n+1}}(Q).
\end{equation}
The argument is precisely analogous to that of section~\ref{sss2.1}
(an almost identical construction gives
essentially a Morita
equivalence $T_{A_{n-1}} \sim T_{A_n} e_n T_{A_n}$, and so on).
Extended concrete examples for the $A_n$ case in particular are given in
section~9.5.2 of reference \cite{Mar}.

It follows (c.f. section~\ref{sss2.4.3} - but in this case
we make use of the symmetry under reversing the order of the chain,
i.e. swapping $e_1$ and $e_n$) that the
generic
Bratteli restriction diagram for the
irreducible representations             
 associated to $T_n(Q) \supset T_{n-1}(Q)$ (with
top line $n=0$) is:
\[
\begin{array}{cccccccccccccccccc}
&1
\\
&\nearrow\nwarrow
\\
1&&1
\\
&\nwarrow\nearrow
\\
&2&&1
\\
&\nearrow\nwarrow&&\nearrow\nwarrow
\\
2&&3&&1
\\
&\nwarrow\hspace{.5cm}&\nearrow\nwarrow&\hspace{.5cm}\nearrow
\\
&5&&4&&1
\\
&\nearrow\nwarrow&&\nearrow\nwarrow&&
\nearrow\nwarrow
\\
5&&9&&5&&1
\\
&\nwarrow\hspace{.5cm}&\nearrow\nwarrow&&
\nearrow\nwarrow&\hspace{.5cm}\nearrow
\\
&14&&14&&6&&1
\\
&\nearrow\nwarrow&&\nearrow\nwarrow&&\nearrow\nwarrow&&
\nearrow\nwarrow
\\
14&&28&&20&&7&&1
\\
&\nwarrow\hspace{.5cm}&\nearrow\nwarrow&&\nearrow\nwarrow&&
\nearrow\nwarrow&\hspace{.5cm}\nearrow
\\
&42&&48&&27&&8&&1
\\
&\nearrow\nwarrow&&\nearrow\nwarrow&&\nearrow\nwarrow&&\nearrow\nwarrow&&
\nearrow\nwarrow
\\
42&&90&&75&&35&&9&&1
\\
&\nwarrow\hspace{.5cm}
&\nearrow\nwarrow&&\nearrow\nwarrow&&\nearrow\nwarrow&&
\nearrow\nwarrow&\hspace{.5cm}\nearrow
\\
&132&&165&&110&&44&&10&&1
\\
&\nearrow\nwarrow&&\nearrow\nwarrow&&\nearrow\nwarrow&&\nearrow\nwarrow&&
\nearrow\nwarrow&&\nearrow\nwarrow
\\
132&&297&&275&&154&&54&&11&&1
\end{array}
\]
and so on.
Recall, for comparison with
the diagram for $P_n(Q)$, and with the diagram in equation (\ref{XX})
 later, that
$T_{2n}(Q) \cong T_{A_{n}}(Q)$, i.e. we jump two lines on this diagram
for every one on the diagram in equation (\ref{XX}).
Also note that here, unlike $P_n(Q)$, there is only a single
representation associated with each node. This is because the
planar 2d lattice allows no possibility of connectivities
crossing without themselves becoming connected (i.e. no
permutations are allowed).

It is straightforward to compute the exceptional structures in this
case (see \cite{Mar}).

We can summarize the answers to the questions posed at the end of the previous
sub section in this case by noting that here the Diagram algebra is isomorphic
to the Full algebra.

\vspace{.3in}

Perhaps the next two simplest cases to consider are
$G=\hat{A}_n$ and the daisy graphs
(tree graphs consisting of spokes radiating from a central node).
They are both extremely illuminating, and we will deal with each
in commensurate detail.

\section{The case $\hat{A}_n$}
Here we want to
address the following questions:

1. What is the structure of $D_{G}(Q)$?

2. What is the structure of $T_{G}(Q)$?

3. What is the content of the TM algebra, i.e. the
cylindrical Potts model
representation?

Everything we do in this section is enthralled by the consequences
for $\hat{A}_n$
of the highly useful
$A_n$ tower relation~(\ref{tower}).
For $A_n$ this is the key to an inductive step
between the structure of the smaller and the
larger algebra.
Here, of course, we can only manage a weak echo
\begin{equation}
\label{Y}
\begin{array}{ccccc}
\underline{n} & \subset & \underline{n+1}
\\
\cup   &                & \cup
\\
\hat{A}_n  &  \not\subset  &  \hat{A}_{n+1}
\\
\cup  &                & \cup
\\
A_n       &   \subset    & A_{n+1}.
\end{array}
\end{equation}
where the vertical inclusions are realized by, for example,
omitting $U_{1n}$ in going from $\hat{A}_n$ to $A_n$.
We are further handicapped by the extra complications of even
the
simplest cases:

\subsection{The case $D_{\hat{A}_n}(Q)$}

This is the quotient
of $T_{\hat{A}_n}(Q)$ appropriate for dichromatic polynomials
(that is to say, the representations appearing in the TMs
for dichromatic polynomials are representations of this algebra
\cite{Mar}), and
so it is reasonable to suppose that it plays an important role in
the closely related Potts models as well (see subsection \ref{The Potts
model}),
 if not in all
physical models.

We would like to know the complete structure of the algebra,
but physically the
 key initial question is: What is the generic asymptotic growth
rate, with $n$, of dimensions of irreducible representations?
The asymptotic growth rate for the $Q$-state Potts
representation is $Q$.  Clearly
if the generic asymptotic growth rate of dimensions of irreducible
representations is greater than the actual asymptotic growth rate
of the model representation
generically containing
those irreducibles then eventually the model representation
must become smaller than the generic
irreducibles and hence a source of non-generic representations.
For axiomatic statistical mechanics
this crossover number should still be 4
here (i.e. as it is for the open boundary case). This is indeed what we find.

Bearing in mind our tower structure let us begin by recalling the
generic
Bratelli restriction diagram for the
irreducible representations in the ordinary open bounded case,
i.e. associated to $T_{A_n}(Q) \supset T_{A_{n-1}}(Q)$. Starting with
top line $n=0$ we have:
\begin{equation}
\label{XX}
\begin{array}{cccccccccccccccccc}
1
\\
\up&\nw
\\
1&&1
\\
\up&\nx&\upp&\nw
\\
2&&3&&1
\\
\up&\nx&\upp&\nx&\upp&\nw
\\
5&&9&&5&&1
\\
\up&\nx&\upp&\nx&\upp&\nx&\upp&\nw
\\
14&&28&&20&&7&&1
\\
\up&\nx&\upp&\nx&\upp&\nx&\upp&\nx&\upp&\nw
\\
42&&90&&75&&35&&9&&1
\\
\up&\nx&\upp&\nx&\upp&\nx&\upp&\nx&\upp&\nx&\upp&\nw
\\
132&&297&&275&&154&&54&&11&&1
\end{array}
{}.
\end{equation}

Since the affine case restricts to the open case by omitting
$U_{1n}$ we know that the               
irreducible representations
must decompose as                       
direct sums of the representations above.
The irreducible representations may be characterised in essentially
the same way
as in the open case (quotienting
by those     
permutations of $t=0$ to $t=T$ lines discussed above,
which in this case are just
cyclic permutations, the number of equivalence classes is the same),
but the dimensions are different. These may be computed in a
number of alternative ways.  The above
dimensions will be lower bounds. The $P_n(Q)/\sim$ dimensions in
section~\ref{sX} will be upper bounds.

We find the following table of dimensions
of irreducibles for $T_{\hat{A}_n}(Q)$
(not a restriction diagram this time -  see equation (\ref{Y})):
\[
\begin{array}{cccccccccccccccccc}
1
\\
1&&1
\\
2&&3&&1
\\
5&&10&&6&&1
\\
14&&35&&28&&8&&1
\\
42&&126&&120&&45&&10&&1
\\
132&&462&&495&&220&&66&&12&&1
\end{array}
\]
and so on.
Note that the numbers in column $i$ (counting from $i=0$ on
the left) give the dimensions, but correspond in general to
more than one inequivalent irreducible, since a semi-direct
product with the cyclic group of order $i$ survives from
the
corresponding product with $\Sigma_i$ in the `mean field' case
($G=\underline{n}$).
For a general graph $G$ our procedure is first to compute the
$P_n(Q)/\sim$ type cases (which symmetrise among the
possible permutations of distinct connectivities), and
then semi-direct product (as in section~\ref{sX}) with
the subgroup of $\Sigma_i$ consistent with the
permutations that can be realised in $G\times\Z$. In this case
only cycles are
available (and all irreducible dimensions thus multiplied by 1).

For example the $S_3(2) \Sigma_\pm$ basis may be written
 \[
\{ (
((121')(32'))\pm((122')(31'))), \;\;
(((232')(11'))\pm((231')(12'))),          \;\;
\]
\[
\;\;
(((22')(131'))\pm((21')(132'))),
           \;\;
(((1)(21')(32'))\pm((1)(22')(31'))),                 \;\;
\;\;
\]
\[ \;\;\;\;\;
(((2)(11')(32'))\pm((2)(12')(31'))), \;\;    \;\;
(((3)(11')(22'))\pm((3)(12')(21')))   \; \}
\]
so  the representation of $1_{12}$ for the $-$ case is
  \[
{\cal R}_{(1^3)} \left( 1_{12} \right)
=
\left( \begin{array}{cccccccc}
1&0&0&0&0&0
\\
0&0&\pm1&0&0&0
\\
0&\pm1&0&0&0&0
\\
0&0&0&0&1&0
\\
0&0&0&1&0&0
\\
0&0&0&0&0&\pm1
\end{array}
\right) .
\]
and is the  identity matrix for the $+$ case.

The dimensions may be computed as follows.
Let $C_{n,j}$ denote the irreducible representation of $T_{A_n}(Q)$
with $j$ distinct lines from $t=0$ to $t=T$;
the representation induced from
$D_{A_n}  
I_j / D_{A_n} I_{j-1} D_{A_n}$
 (i.e. the representation
in the $n^{th}$ row and $j^{th}$ column of the diagram above, in each
case counting from zero on the top, resp. left). Let $\hat{C}_{n,j}$
denote the corresponding representation of $T_{\hat{A}_n}(Q)$.
Then
\begin{pr}
\label{dims}
\begin{equation}
\label{dim0}
\hat{C}_{n,0} \downarrow_{T_{A_n}} = C_{n,0}
\end{equation}
\begin{equation}
\label{dim1}
\hat{C}_{n,1} \downarrow_{T_{A_n}} = \sum_{m=0,1,2,...} C_{n,2m+1}
\end{equation}
and for $k=2,3,4,...n$
\begin{equation}
\label{dim2}
\hat{C}_{n,k} \downarrow_{T_{A_n}} = \sum_{m=0,1,2,...,n-k} C_{n,k+m}
\end{equation}
\end{pr}
{\em Proof:}
The dimension $C_{n,0}$ is given by the number of elements of
$S_n=S_n(0)$ such that $i \sim j$, $k \sim l$  and $i<k<j$ implies
that $i<l<j$ \cite{Mar}.
The first equation comes from noting that the same constraint applies
for $\hat{C}_{n,0}$.

The dimension $C_{n,1}$ is given by the number of elements of
$S_n(1)$ such that $i \sim j$, $k \sim l$  and $i<k<j$ implies
that $i<l<j$; and such that $i \sim j$ implies no primed part
contains $k$ such that $i<k<j$.
The second equation comes from noting that the
difference
between this and
 $\hat{C}_{n,1}$ is elements in
the basis for the latter where a connection $i \sim j$
does not prevent $k$ such that $i<k<j$ appearing in a primed part
(we write $k \sim \infty$).
Consider these elements and modify them by
cutting the line from $i$ to $j$ and attaching each loose end
separately to $\infty$ (i.e. put these two in separate
primed parts).
That this is the correct move follows from the action
of the generators of $T_{A_n}(Q)$ on the basis states
of $\hat{C}_{n,1}$.
These elements
include the whole of $C_{n,3}$
plus any for which there was
originally a further $i_2 \sim j_2$
and intermediate $k_2 \sim \infty$.
Cutting again, and so on, we get the
required result.          

For the third equation, note that $\hat{C}_{n,k} \supset C_{n,k}$.
The difference is elements where a line  to $\infty$ is `isolated'
from $\infty$ in $C_{n,k}$ as above.
The innermost of these isolating lines may have its own line to $\infty$ also
(subsequent ones may not, even in $\hat{C}_{n,k}$).
Thus we get either zero or one lines to $\infty$ becoming two
lines to $\infty$ by cutting at the problem point.
The new element is either in $C_{n,k+1}$ or $C_{n,k+2}$ or
still illegal (but now with
either $k+1$ or $k+2$ lines to
$\infty$). It is straightforward to check that all of
these two bases is generated in this way. Cutting again
on the illegal elements we may iterate to obtain the required results. QED.

{}From proposition \ref{dims} the dimensions of representations in $A$ and
$\hat{A}$ cases are linearly related. Therefore the asymptotic growth rate is
equal to 4
in both cases.

\subsection{The Full TL case $T_{\hat{A}_n}(Q)$: explicit example for $n=2$}

\subsubsection {Generalities}

Consider the unital associative algebra over $\C$ defined by generators
\[
1,U_{1.},U_{2.},U_{12},U_{21}
\]
and relations~\ref{r4}-\ref{r6}.

This is a `degenerate' case of $T_G(Q)$, i.e.
$G= \hat{A}_2$
(from the diagrammatic point of view we
have two bonds between 1 and 2, so $U_{12} \ne U_{21}$).
 However, it is 
the
simplest interesting case to consider.

It is useful to get an idea of
the problem by simply begining to list linearly independent words in
this algebra by length:
\[
1
\]
\[
U_{1.}, \;\; U_{2.}, \;\; U_{12}, \;\; U_{21}
\]
\[
U_{1.} U_{2.},  \;\; U_{1.} U_{12}, \;\; U_{1.} U_{21}, \;\; U_{21} U_{12},
\;\;
  U_{2.} U_{12},  \;\; U_{2.} U_{21}, \;\; U_{12} U_{1.}, \;\;
U_{12}U_{2.}, \;\; U_{21}U_{1.}, \;\;  U_{21} U_{2.}
\]
\[
U_{1.} U_{2.} U_{12}, \;\;
U_{1.} U_{12} U_{2.}, \;\;
 U_{1.} U_{21} U_{2.}, \;\;
U_{1.}U_{21} U_{12}, \;\;
  U_{2.} U_{12}U_{1.}, \;\;
 U_{1.} U_{2.} U_{21},
\hspace{1in}
\]
\[
\hspace{1in}
U_{12} U_{1.} U_{2.}, \;\;
U_{2.}U_{12}U_{21},  \;\;
U_{21}U_{1.} U_{2.},
\;\; U_{21} U_{2.} U_{12}, \;\;
U_{21}U_{1.}U_{12}, \;\;
U_{12}U_{1.}U_{21}
\]
...and so on (we have given the
complete list up to length 3).
 Clearly we need to be able to get some control of this
explosion of elements. Our
treatment of $P_n(Q)$
tells us
how to do this.
There we were able to filter
the algebra through a sequence of
double sided ideals generated by
certain idempotents. Since that was a
quotient algebra of this, we still have these
elements
as candidates for primitive idempotents.
Let us look at the left sided ideal generated by
(unnormalized) idempotent $U_{1.}U_{2.}$.
Organising the words by length we have the following structure
\begin{equation}
\begin{array}{ccccc}
&U_{1.} U_{2.}
\\
\\
U_{12} U_{1.} U_{2.}
&&
U_{21} U_{1.} U_{2.}
\\
&U_{12} U_{21} U_{1.} U_{2.}
\\
\\
...&&...
\\
\\
&\left( U_{12} U_{21}U_{1.} U_{2.} \right)^k &
\\
\\
U_{1.} \left( U_{12} U_{21}U_{1.} U_{2.} \right)^k
&& U_{2.} \left( U_{12} U_{21}U_{1.} U_{2.} \right)^k
\\
\\
&\!\!\!\!\! U_{1.}U_{2.} \left( U_{12} U_{21}U_{1.} U_{2.} \right)^k
\!\!\!\!\!&
\\
\\
U_{12}U_{1.}U_{2.} \left( U_{12} U_{21}U_{1.} U_{2.} \right)^k \!\!\!\!\!
&& \!\!\!\!\! U_{21} U_{1.} U_{2.} \left( U_{12} U_{21}U_{1.} U_{2.} \right)^k
\\
\\
&...
\end{array}\label{ugly}
\end{equation}

The ideal generated by $U_{1.}U_{2.}
\left( U_{12} U_{21}U_{1.} U_{2.} \right)^k$ (any natural number $k$)
is a proper invariant subspace (see proposition \ref{godknows}).
  There is an injective map from
$TU_{1.}U_{2.}$ into $TU_{1.}U_{2.}\left( U_{12} U_{21}U_{1.} U_{2.}
\right)^k$,
so the full ideal $TU_{1.}U_{2.}$
 is isomorphic to a direct product of the $k=0$
sector with the natural numbers under addition,
$({\bf N\!I},+)$ - which is abelian.
A general source
of irreducible representations is thus the quotient of $TU_{1.}U_{2.}$
by
\begin{equation}
U_{1.}U_{2.} \left( U_{12} U_{21}U_{1.} U_{2.} \right)
 = \alpha U_{1.}U_{2.}\label{quotient}
\end{equation}
where $\alpha \in \C$ corresponds to the irrep. of $({\bf N\!I},+)$
given by $R(1)=\alpha$.

These quotients have a symmetric inner product $<>$ defined by
\begin{equation}
AB^T = <AB> U_{1.}U_{2.}\label{henri}
\end{equation}
with Gram matrix, in the basis
\begin{equation}
\{
U_{1.}U_{2.}, \;\;
U_{12} U_{1.}U_{2.},\;\;
U_{21} U_{1.}U_{2.},\;\;
U_{12}U_{21} U_{1.}U_{2.}, \;\;
U_{1.} U_{12}U_{21}U_{1.}U_{2.}, \;\;
U_{2.} U_{12}U_{21}U_{1.}U_{2.}
\}\label{basis}
\end{equation}

that reads
\[
\hbox{Gr}_{n=2,i=0}=\left(
\begin{array}{ccccccccc}
Q&\sqrt{Q}&\sqrt{Q}&\alpha&\alpha \sqrt{Q}& \alpha \sqrt{Q}
\\
\sqrt{Q}&Q&\alpha &\alpha \sqrt{Q}&\alpha & \alpha
\\
\sqrt{Q}&\alpha &Q&\alpha \sqrt{Q}&\alpha & \alpha
\\
\alpha &\alpha \sqrt{Q}&\alpha \sqrt{Q}&\alpha Q&\alpha \sqrt{Q}&
\alpha \sqrt{Q}
\\
\alpha \sqrt{Q}&\alpha &\alpha &\alpha \sqrt{Q}&\alpha Q& \alpha^2
\\
\alpha \sqrt{Q} &\alpha&\alpha &\alpha \sqrt{Q}&\alpha^2& \alpha Q
\end{array} \right)
\]
The latter has for determinant
\begin{equation}
\hbox{det Gr}_{2,0}=\alpha^3 (\alpha - Q)^4 ((Q-2)^2 - \alpha)\label{deter}
\end{equation}

With the relation (\ref{quotient}) $U_{1.}U_{2.}/Q$ becomes a primitive
idempotent, so the representation is indecomposable. Taking also into account
the Gram matrix it is thus generically irreducible.

We still have a representation in terms of boundary diagrams
 but
now
lines
are non-contractible non-overlaping segments on,
and
begining and ending on the   boundaries
of,
the twice punctured sphere without any seam:
figure~\ref{U12pic}.
\begin{figure}
\begin{picture}(100,100)(-100,30)
\put(65,100){\oval(10,20)[b]}
\put(45,100){\oval(10,20)[b]}
\put(45,40){\oval(10,20)[t]}
\put(65,40){\oval(10,20)[t]}
\put(55,120){\oval(90,160)[b]}
\put(55,120){\oval(90,40)}
\end{picture}
\caption{Boundary
diagram for $U_{1.} U_{2.}$ in $D_{\hat{A}_2}(Q)$ drawn on a cylinder.
 \label{U12pic}}
\end{figure}
It follows from the structure in (\ref{ugly})
 that this is a faithful representation.
The ideal $TU_{1.}U_{2.}$ is  infinite dimensional
. We can write down a sequence of quotients which
exhibit the invariant subspaces indicated by the determinant (\ref{deter}).
With the above topological picture in  mind,
the first trial
quotient
with which we might hope to collapse the ideal
into a non-trivial finite dimensional one, is of the form
\[
(e1+aU_{12}+bU_{21}) \; U_{1.}U_{2.} =0
\]
($e,a,b$ scalars) but then
\[
U_{1.}
(e1+aU_{12}+bU_{21}) \; U_{1.}U_{2.} =
(\sqrt{Q} e  + a +b)
\; U_{1.}U_{2.} =0
\]
and
\[
U_{12} U_{21}
(e1+aU_{12}+bU_{21}) \; U_{1.}U_{2.} =
( e  + \sqrt{Q}( a +b)) U_{12} U_{21}
\; U_{1.}U_{2.} =0
\]
so either $Q=1$ (an exceptional case,
which we will discount for the moment),
 or $e=0$ and $a=-b$ ($=-1$, without
loss of generality).
Diagramatically we have
equivalence between the pictures of figure~\ref{U12pi} and
figure~\ref{U12},
\begin{figure}
\begin{picture}(100,100)(-100,30)
\put(65,100){\oval(10,20)[b]}
\put(45,100){\oval(10,20)[b]}
\put(45,60){\oval(10,20)[t]}
\put(65,60){\oval(10,20)[t]}
\put(55,60){\oval(10,10)[b]}
\put(55,40){\oval(10,10)[t]}
\put(40,40){\line(0,1){20}}
\put(70,40){\line(0,1){20}}
\put(55,120){\oval(90,160)[b]}
\put(55,120){\oval(90,40)}
\end{picture}
\caption{Boundary
diagram for $U_{1.} U_{2.}U_{12}$ in $D_{\hat{A}_2}(Q)$. \label{U12pi}}
\end{figure}
\begin{figure}
\begin{picture}(100,100)(-100,30)
\put(65,100){\oval(10,20)[b]}
\put(45,100){\oval(10,20)[b]}
\put(45,60){\oval(10,20)[t]}
\put(65,60){\oval(10,20)[t]}
\put(35,60){\oval(10,10)[br]}
\put(35,40){\oval(10,10)[tr]}
\put(35,70){\oval(50,30)[bl]}
\put(35,60){\oval(50,30)[bl]}
\put(75,60){\oval(10,10)[bl]}
\put(75,40){\oval(10,10)[tl]}
\put(75,70){\oval(50,30)[br]}
\put(75,60){\oval(50,30)[br]}
\put(50,40){\line(0,1){20}}
\put(60,40){\line(0,1){20}}
\put(55,120){\oval(90,160)[b]}
\put(55,120){\oval(90,40)}
\end{picture}
\caption{Boundary
diagram for $U_{1.} U_{2.}U_{21}$ in $D_{\hat{A}_2}(Q)$. \label{U12}}
\end{figure}
i.e. we have legalized this topologically
non-trivial move.
Altogether the prospective
quotient ideal is spanned by
\begin{equation}
\label{Xx}
\{
U_{1.} U_{2.},  \;\; U_{12}U_{1.}U_{2.}
\}
\end{equation}
and $U_{1.}U_{2.}$ is a primitive idempotent (so the ideal
 is
indecomposable, and generically
simple).
It is easy to see that this module is isomorphic to the 2 dimensional
representation of the diagram algebra, which is, in turn, the
one induced by braid translation
\[
U_{1n}=\left( \prod_{i=1}^{n-1} g_{i.}g_{ii+1} \right)^{-1}
U_{1.}
\left( \prod_{i=1}^{n-1} g_{i.}g_{ii+1} \right)
\]
($g=1-q^{\pm 1}U$ where $q+q^{-1} =\sqrt{Q}$)
from the $A_n$ case (see figure \ref{bt}).
\begin{figure}
\vspace{4cm}\label{bt}
\end{figure}

There are many other possible quotients which we may try. For example,
\[
U_{12}U_{21}U_{1.}U_{2.}= k \; U_{1.}U_{2.}
\]
where $k$ is a scalar. In this case
\[
U_{21} U_{12} \;\; U_{12}U_{21}U_{1.}U_{2.}=
Q \;\; U_{21} U_{12} \;\; U_{1.}U_{2.} =
U_{21} U_{12} \;\; k \; U_{1.}U_{2.}
\]
so $k=Q$, and
\[
U_{1.} \;\; U_{21} \;\; U_{12}U_{21}U_{1.}U_{2.}=
U_{1.} \;\; U_{21} \;\; k \; U_{1.}U_{2.}
=
k \;\; U_{1.} U_{2.}
=
Q k U_{1.} U_{2.}
\]
so $Q=1$. For the moment we will disregard such exceptional $Q$ cases.
As is probably already apparent, the way to get generic cases is to make
 quotients
corresponding to introducing topological moves (as  equation (\ref{Xx})).
There are various options corresponding to simply dragging a
line across a puncture, which we will leave to the interested
reader. Those which give rise to representations in which a new
parameter actually survives unconstrained
 are characterized in the next section:

\subsubsection{Free non-contractible loops}

By reference to the boundary diagram picture we note that a natural class
of (finite dimensional) quotients to consider is
\begin{equation}
U_{1.} U_{2.}
( U_{12} U_{21}
U_{1.} U_{2.} )^{p_1}
=
\alpha(p_1,p_2) \;\;
U_{1.} U_{2.}
( U_{12} U_{21}
U_{1.} U_{2.} )^{p_2}\label{p1p2}.
\end{equation}
This corresponds
diagrammaticaly
to establishing a rule in which $2p_1$ free non-contractible
loops (see figure~\ref{U12p})
may be replaced by $2p_2 < 2p_1$ loops at cost of an overall
factor $\alpha(p_1,p_2)$.
\begin{figure}
\begin{picture}(100,100)(-100,30)
\put(65,100){\oval(10,20)[b]}
\put(45,100){\oval(10,20)[b]}
\put(45,40){\oval(10,20)[t]}
\put(65,40){\oval(10,20)[t]}
\put(55,120){\oval(90,160)[b]}
\put(55,120){\oval(90,40)}
\multiput(55,110)(0,5){4}{\oval(90,100)[b]}
\end{picture}
\caption{Boundary
diagram for $p_1 =2$ in $D_{\hat{A}_2}(Q)$. \label{U12p}}
\end{figure}
The dimension is $6 p_1$. However the representation implied by (\ref{p1p2})
reduces to a direct sum of cases $p_1=1,p_2=0$ (\ref{quotient}) for various
choices of $\alpha$. Consider for instance the case $p_2=0$. Call $X$ any
element of the basis (\ref{basis}). The subspace spanned by ("topological"
Fourier transform)
\[
|X,k>=\sum_{p=0}^{p_1-1} \alpha(p_1)^{\frac{pk}{p_1}}\
X\left(U_{12}U_{21}U_{1.}U_{2.}\right)^p
\]
is invariant for each $k=0,\ldots,p_1-1$. The relation (\ref{quotient}) holds
with $\alpha=\alpha(p_1)^{\frac{k}{p_1}}$. If $p_2\neq 0$ the sector with
 $2p_2$  loops is an invariant subspace isomorphic to the case $p_2=0$.
The quotient has $\alpha=0$.

Thus the only case to consider is the aforementioned (\ref{quotient})
\[
U_{1.} U_{2.}
U_{12} U_{21}
U_{1.} U_{2.}
=
\alpha \;\;
U_{1.} U_{2.}
\]
with $\alpha$ a scalar.
A basis for the full ideal with the above quotient is  given by (\ref{basis})
i.e.
this
produces a 6 dimensional algebra (without unit).
Here
 $U_{1.}U_{2.}/Q$
{\em is}
a primitive idempotent, so the corresponding representation is
indecomposable.
In our basis we have (with all omitted entries zero)
\[
R_{\alpha} \left( U_{1.}  \right)
=
\left(
\begin{array}{ccccccc}
\sqrt{Q}
\\
1&0
\\
1&0&0
\\
0&0&0&0&1
\\
0&0&0&0&\sqrt{Q}
\\
\alpha&0&0&0&0&0
\end{array}
\right)
\hspace{.2in}
R_{\alpha} \left( U_{2.} \right)
=
\left(
\begin{array}{ccccccc}
\sqrt{Q}
\\
1&0
\\
1&0&0
\\
0&0&0&0&0&1
\\
\alpha&0&0&0&
\\
0&0&0&0&0&\sqrt{Q}
\end{array}
\right)
\]
\[
R_{\alpha} \left( U_{12} \right)
=
\left(
\begin{array}{ccccccc}
0&1
\\
0&\sqrt{Q}&
\\
0&0&0&1
\\
0&0&0&\sqrt{Q}&
\\
0&0&0&1&0
\\
0&0&0&1&0&0
\end{array}
\right)
\hspace{.2in}
R_{\alpha} \left( U_{21} \right)
=
\left(
\begin{array}{ccccccc}
0&0&1
\\
0&0&0&1
\\
0&0&\sqrt{Q}&
\\
0&0&0&\sqrt{Q}&
\\
0&0&0&1&0
\\
0&0&0&1&0&0
\end{array}
\right).
\]

%
\begin{pr}
The algebra $T_{\hat{A}_2}(Q)$
is infinite dimensional.  \label{godknows}
\end{pr}

{\em Proof}:
We  prove that the above set of representations characterized by
the indeterminate $\alpha$ are mutually inequivalent.
This follows by direct computation of
\[
trace \left( R_{\alpha} \left( U_{12}U_{21}U_{1.}U_{2.} \right) \right)
= \alpha
{}.
\]
QED

A straightforward generalisation of this proof 
serves to
show that the $\hat{A}_n$ algebra is infinite dimensional for all $n>1$.

If we want to examine the exceptional structure
of the present set of representations with respect to $\alpha$
then one suitable basis is
\[
\{
U_{1.}U_{2.},\;\;
(U_{12} + U_{21})U_{1.}U_{2.},\;\;
(U_{12} - U_{21})U_{1.}U_{2.},\;\;
(\sqrt{Q} U_{12} - U_{12} U_{21})U_{1.}U_{2.},
\]
\[
\;\;
(\sqrt{Q}  - U_{1.} U_{12} U_{21})U_{1.}U_{2.},\;\;
(\sqrt{Q}  - U_{2.} U_{12} U_{21})U_{1.}U_{2.}
\}
\]
in terms of which we have
\[
U_{1.}
=
\left(
\begin{array}{ccccccc}
\sqrt{Q}
\\
2&0
\\
0&0&0
\\
0&0&0&0&1
\\
0&0&0&0&\sqrt{Q}
\\
Q-\alpha&0&0&0&0&0
\end{array}
\right)
\hspace{.2in}
U_{2.}
=
\left(
\begin{array}{ccccccc}
\sqrt{Q}
\\
2&0
\\
0&0&0
\\
0&0&0&0&0&1
\\
Q-\alpha&0&0&0&
\\
0&0&0&0&0&\sqrt{Q}
\end{array}
\right)
\]
\[
U_{12}
=
\left(
\begin{array}{ccccccc}
0&1/2&1/2
\\
0&\sqrt{Q}&\sqrt{Q}&-1
\\
0&0&0&1
\\
0&0&0&\sqrt{Q}&
\\
0&0&0&1&0
\\
0&0&0&1&0&0
\end{array}
\right)
\hspace{.2in}
U_{21}
=
\left(
\begin{array}{ccccccc}
0&1/2&-1/2
\\
0&\sqrt{Q}&0&-1
\\
0&0&\sqrt{Q}&1
\\
0&0&0&0&
\\
0&0&-\sqrt{Q}&1&0
\\
0&0&-\sqrt{Q}&1&0&0
\end{array}
\right).
\]
We note that when $\alpha=Q$
(c.f. the Gram matrix)
there is a two dimensional invariant subspace
corresponding to the quotient
\[
U_{12}U_{1.}U_{2.}
=
U_{21}U_{1.}U_{2.}
\]
which we have already identified with the diagram representation
(equation (\ref{Xx})).

Note that the left sided ideal generated by $U_{12}U_{21}$ will be
`dual' to this one (i.e. isomorphic up to the usual duality
transformation).

\subsubsection{The case $TU_{1.}/TI_0T$}
 The word structure of the left sided
ideal generated by $U_{1.}$ (quotienting by $TU_{1.}U_{2.}T$) is
\[
\begin{array}{ccccccccccccccccccc}
 & & & U_{1.}
\\
 & &U_{12}U_{1.}&&U_{21}U_{1.}
\\
&U_{2.}U_{12}U_{1.}&&U_{12}U_{21}U_{1.}&&U_{2.} U_{21} U_{1.}
\\
U_{21} U_{2.}U_{12}U_{1.}&&U_{1.}U_{12}U_{21}U_{1.} && U_{2.}U_{12}U_{21}U_{1.}
                                    &&U_{12} U_{2.} U_{21} U_{1.}
\end{array}
\]
and so on.
By the same argument as before, the
appropriate quotient here is
\begin{equation}
U_{1.}  U_{21} U_2U_{12}U_{1.} = \alpha_1 U_{1.}\label{quotient1}
\end{equation}
and
\begin{equation}
U_{1.} U_{12} U_{21} U_{1.}=zU_{1.}
\end{equation}
Notice that (\ref{quotient1}) implies
\begin{equation}
U_{1.}U_{12}U_{2.}U_{21}U_{1.}=\alpha_1^{-1}U_{1.}
\end{equation}
The only consistent choices are $z=0$ or $z=\sqrt{Q},\alpha_1=1$.
This
then reduces to a four dimensional ideal. Once again there is
a symmetric inner product, with Gram matrix in the basis $\left\{
U_{1.},U_{1.}U_{12},
U_{1.}U_{12}U_{2.},U_{1.}U_{12}U_{2.}U_{21}\right\}$,
\[
\hbox{Gr}_{2,1}=\left(
\begin{array}{ccccc}
\sqrt{Q}&1&0&\alpha_1
\\
1&\sqrt{Q}&1&z
\\
0&1&\sqrt{Q}&1
\\
\alpha_1^{-1} &z&1&\sqrt{Q}
\end{array}   \right)
\]
with determinant
\begin{equation}
(-(\alpha_1+\alpha_1^{-1})+Q^2-4Q+2)
\end{equation}
in case $z=0$. In case $z=\sqrt{Q}$ the irreducible quotient
is isomorphic to $\hat{C}_{n,1}$.

In order to see what is going on it is useful to
look again at  figure~\ref{U12ppp}: a boundary gets a weight $\alpha_1^{\pm 1}$
when going around the cylinder.
\begin{figure}
\begin{picture}(100,100)(-100,30)
\put(45,100){\oval(10,20)[b]}
\put(60,90){\line(0,1){10}}
\put(70,90){\line(0,1){10}}
\put(45,40){\oval(10,20)[t]}
\put(70,40){\line(0,1){10}}
\put(60,40){\line(0,1){10}}
\put(55,120){\oval(90,160)[b]}
\put(55,120){\oval(90,40)}
\put(60,50){\line(-2,1){40}}
\put(70,50){\line(-2,1){50}}
\put(60,90){\line(2,-1){30}}
\put(70,90){\line(2,-1){20}}
\end{picture}
\caption{Boundary
diagram for $U_1 U_{12} U_2 U_{21} U_1$.
 \label{U12ppp}}
\end{figure}

\subsection{The Full case $T_{\hat{A}_n}(Q)$}

To get further insight we shall use   the analysis of the vertex model
carried out  in \cite{PS}. Introduce the space ${\C}^{2n}$ of
dimension $2^{2n}$ considered as tensor product of $2n$ fundamental
representations of the quantum algebra $U_qsl(2)$ (with generators $X_{\pm},H$)
{}.
 Set as usual $\sqrt{Q}=q+q^{-1}$ and consider
 the matrix representation of $T_{\hat{A}_n}(Q)$
where $U_i$ ($U_{i,i+1}$)  acts as identity everywhere except
 in the $(2i-1)^{\mbox{th}}$ and
$(2i)^{\mbox{th}}$ ($(2i)^{\mbox{th}}$ and
$(2i+1)^{\mbox{th}}$) copies of ${\C}^2$ where it acts as the  $4\times 4$
matrix,
\[
U=\left(\begin{array}{cccc}
0&0&0&0\\
0&q^{-1}&-1&0\\
0&-1&q&0\\
0&0&0&0
\end{array}\right)
\]
while the special boundary element $U_{1n}$  acts between the first and last
copies of ${\C}^2$ as
\[
U=\left(\begin{array}{cccc}
0&0&0&0\\
0&q^{-1}&-x&0\\
0&-x^{-1}&q&0\\
0&0&0&0
\end{array}\right)
\]
where $x$ is a complex parameter.
For each value of the weight $h=0,\pm 1,\ldots,\pm n$ this provides a
representation of $T_{\hat{A}_n}(Q)$ with continuous parameter $x$. We
shall refer to this representation as $R_n\left(\begin{array}{c}
x\\
h
\end{array}\right)$. It has for dimension the binomial coefficient
$C^{n-h}_{2n}$.
The cases $\pm h$ are isomorphic and we
restrict without loss of generality to the study of $h\geq 0$.

\subsubsection{The case $TI_0$}

The same analysis  as $n=2$ works for any $n$,
i.e. all irreducible representations may be found by putting
\begin{equation}
\left(U_{1.}U_{2.}...U_{n.}\right)  U_{12}U_{23}...U_{n1}
\left(U_{1.}U_{2.}...U_{n.}\right)=\alpha
\left(U_{1.}U_{2.}...U_{n.}\right)\label{genquo}
\end{equation}

\begin{pr}
The dimension of this sector is ${\cal C}_n =(n+1).C_{n,0}=C_{2n}^n$.
\end{pr}

 This is because
in addition to the $C_{n,0}$ basic connectivities in the
$i=0$ (zero propagating connections) sector we get a new basis state
from each one of these by picking any of the $n$ sites and
moving it round a closed non-contractible path.

The idempotent $I_O$ is primitive with the quotient (\ref{genquo}) so $TI_0$
modulo (\ref{genquo}) is  indecomposable. It has a natural basis of words in
the generators generalizing (\ref{basis}). The representation induced from this
basis we will call ${\cal T}_n\left(\begin{array}{c}
x\\
0
\end{array}\right)$.

There is a simple intertwiner between ${\cal T}_n$
  and the vertex model $R_n\left(\begin{array}{c}
x\\
0
\end{array}\right)$ where
\[
\alpha=\left(x^{1/2}+x^{-1/2}\right)^2.
\]
 Associate with each boundary diagram a state in the $h=0$
sector of the vertex model basis by giving all possible  orientations
 to boundaries with the weights

\begin{equation}
\wedge\!\!\!\!\bigcap\!\!\!\!\vee=\wedge\!\!\!\!\bigcup\!\!\!\!\vee=q^{-1/2},
\ \vee\!\!\!\!\bigcap\!\!\!\!\wedge=\vee\!\!\!\!\bigcup\!\!\!\!\wedge
=q^{1/2}\label{bulk}
\end{equation}
and
\begin{eqnarray}
\rfloor^{\;\!\!\!\!\!\vee}\ldots\lfloor^{\:\!\!\!\!\!\wedge}=q^{-1/2}x^{1/2},\
\rceil_{\;\!\!\!\!\!\vee}\ldots\lceil_{\;\!\!\!\!\!\wedge}=q^{-1/2}x^{-1/2}
\nonumber\\
\rfloor^{\;\!\!\!\!\!\wedge}\ldots\lfloor^{\;\!\!\!\!\!\vee}=-q^{1/2}x^{-1/2},\
\rceil_{\;\!\!\!\!\!\wedge}\ldots\lceil_{\;\!\!\!\!\!\vee}=
-q^{1/2}x^{1/2}\label{bdry}
\end{eqnarray}

where up (down) arrow indicates $h=\pm 1/2$. This is best illustrated by an
example. To $U_{12}U_{21}U_{1.}U_{2.}$ for $n=2$ is associated for instance
\[
-q^{1/2}x^{-1/2}|+>\otimes\left(q^{-1/2}|+->-q^{1/2}|-+>\right)\otimes |->
\]
\[
+ q^{-1/2}x^{1/2}|->\otimes \left(q^{-1/2}|+->-q^{1/2}|-+>\right)\otimes |+>
\]
The idempotent $U_{1.}\ldots U_{n.}/Q^{n/2}$
projects the sector $h=0$ of the vertex model onto a one dimensional subspace.
The intertwiner is then obtained by acting on this subspace with the
representations  $R_n\left(\begin{array}{c}
x\\
0
\end{array}\right)$. This intertwiner is $x$-generically invertible as can be
shown by noticing that in every row and column there is exactly one occurence
of the highest power of $q$ (or $q^{-1}$).

The use of the vertex model is to determine a minimal set of invariant
subspaces over the non generic values of $x$. We will then saturate the bound
by reference to $\hbox{Gr}_{n,0}$. It is not convenient to work entirely in
${\cal T}_n$ as the $U_qsl(2)$ action  is not manifest  there.

Commutation with $U_qsl(2)$ is not possible in $\hat{A}_n$ due to
boundary effects, but for non generic values of the parameter $x,\ x=q^{\pm
2k}$
($k$ positive integer) there is a commutative diagram  as the $h=0$ case of
\begin{equation}
\begin{array}{cccccccccccccc}
R_n\left(\begin{array}{c}
x=q^{2k}\\
h
\end{array}\right)&\stackrel{X_+^{k-h}}{\longrightarrow}&
R_n\left(\begin{array}{c}
x=q^{2h}\\
k
\end{array}\right)\\
\downarrow&&\downarrow\\
R_n\left(\begin{array}{c}
x=q^{2k}\\
h
\end{array}\right)&\stackrel{X_+^{k-h}}{\longrightarrow}&
R_n\left(\begin{array}{c}
x=q^{2h}\\
k
\end{array}\right)\\
\end{array}\label{comdiag}
\end{equation}
($k>h$) where $X_+$ is the raising operator in $U_qsl(2)$ \cite{PS}.
 This allows us
to identify an invariant subspace  $\rho_n\left(\begin{array}{c}
k\\
0
\end{array}\right)=\hbox{Ker}X_+^{k}$. A similar result holds by replacing $q$
with $q^{-1}$. Its dimension is $C_{2n}^n-C_{2n}^{n-k}$.
 We note that the intertwiner is singular at these
values.

In addition to this, another case that was overlooked in \cite{PS}
 is $\alpha=0, x=-1$
. Using the symmetry of the $TL$ algebra under $q\rightarrow -q$ one can
establish in that case the existence of a long commutative diagram where the
morphisms  are given by generators of $U_isl(2)$. This  implies that there
is an invariant subspace $\rho'_n$ of dimension
$C_{2n}^n-C_{2n}^{n-1}+C_{2n}^{n-2}\ldots=\frac{1}{2}C_{2n}^n$.

\begin{pr}
The determinant of the Gram matrix
has a factor $\alpha^{{\cal C}_n/2}$
and is overall order $2^{2n-1}$ in $\alpha$.
\end{pr}

Proof. The factor arises since half the basis states {\em begin} with one non
contractible loop (and $AB^T$ has an even number).
An upper bound on the number of closed loops formed by the  $AB^T$
contribution to a single determinant factor
(\ref{henri}) is
 the total number of occurences of $U_{1n}$ in the $TI_0$ basis (since we
cannot build a loop without the periodic closure). In  the $TI_0$ basis there
are $C_{2n}^{2w-1+n}-C_{2n}^{2w+1+n}$ words with $w$ factors of $U_{1n}$, but
\[
\sum_{w=1,2,\ldots}w\left(C_{2n}^{2w-1+n}-C_{2n}^{2w+1+n}\right)=2^{2n-1}
\]

On the other hand the complement
of an invariant subspace contributes a zero of order its dimension to
$\hbox{det Gr}$. The sum of
dimensions of the complements of
the  invariant subspaces we have so far identified in $R_n$ is
\[
\sum_{i=1}^{n}C_{2n}^{n-i}+\frac{1}{2}C_{2n}^{n}=\frac{1}{2}2^{2n}
\]
where we used
\[
(1+1)^{2n}=\sum_{i=0}^{2n}C_{2n}^{i}=C_{2n}^n+2\sum_{i=1}^nC_{2n}^{n-i}.
\]
Therefore the bound is already saturated and this determinant is
\begin{equation}
\hbox{det
Gr}_{n,0}\propto\alpha^{\frac{1}{2}C_{2n}^n}\prod_{i=1}^{n}\left(\alpha-P_i^2
\right)^{C_{2n}^{n-i}}\label{gramdet}
\end{equation}

where we introduced
\[
P_k=q^k+q^{-k}
\]
The $P_k$ are polynomials in $\sqrt{Q}$ determined by the induction
\[
P_k=\sqrt{Q}P_{k-1}-P_{k-2},\ P_0=2,\ P_1=\sqrt{Q}
\]

We have therefore proven
\begin{pr}
The representation $R_n\left(\begin{array}{c}
x\\
0
\end{array}\right)$ has dimension $C_{2n}^{n}$ and
is irreducible except at $x=q^{\pm 2k}$ $k=1,\ldots, n$ and
$x=-1$. When $x=q^{\pm 2k}$ it contains  $\rho_n\left(\begin{array}{c}
k\\
0
\end{array}\right)$ with dimension $C_{2n}^n-C_{2n}^{n-k}$
as an irreducible component. When $x=-1$ it contains
$\rho'_n$ with dimension $\frac{1}{2}C_{2n}^n$
\end{pr}

It is easy to show that the particular $\rho_n\left(\begin{array}{c}
1\\
0
\end{array}\right)$ representation is isomorphic with the one induced from
$C_{n,0}$ by braid translation.

\subsubsection{The case $TI_h/TI_{h-1}T$}

All irreducible representations may be found by introducing the quotient
relation constructed as follows. Take the word $I_h$
( here $I_h=\prod_{i=1}^{n-h}\left(U_{i.}/\sqrt{Q}\right)$)
and rotate the top once
around the  cylinder clockwise holding the bottom fixed. Equate this
 new word with
$\alpha_h I_h$. The same result with $\alpha_h^{-1}$ holds then
for counterclockwise
rotation. For instance if $h=1$
\begin{eqnarray}
\left(U_{1.}\ldots U_{(n-1).}\right)U_{n1}\left(U_{2.}\ldots U_{n.}\right)
U_{12}\ldots U_{n-1,n}
\left(U_{1.}\ldots U_{(n-1).}\right)=\alpha_1 \left(U_{1.}\ldots
U_{(n-1).}\right)\nonumber\\
\left(U_{1.}\ldots U_{(n-1).}\right)U_{12}\ldots U_{n-1,n}\left(U_{2.}\ldots
U_{n.}\right)U_{n1}\left(U_{1.}\ldots U_{(n-1).}\right)
=\alpha_1^{-1} \left(U_{1.}\ldots U_{(n-1).}\right)\label{quotientsunday}
\end{eqnarray}

In the $h=1$ case this does not give a finite dimensional representation. We
need the additional relation
\begin{equation}
\left(U_{1.}U_{2.}\ldots U_{(n-1).}\right)U_{12}U_{23}\ldots U_{n1}\left(U_{1.}
U_{2.}\ldots
U_{(n-1).}\right)=z\left(U_{1.}U_{2.}\ldots U_{(n-1).}
\right)\label{crossclusters}
\end{equation}
This is because a single connectivity can pass through a non contractible
closed loop, while multiple connectivities cannot.
As in the $n=2$ case, either $z=\sqrt{Q},\alpha_1=1$ in which case this is just
$\hat{C}_{n,1}$, or $z=0$ and $\alpha_1$ is unconstrained. Putting $z=0$ brings
$h=1$ into line with other $h>0$ for which this issue does not arise.

The left sided ideal $TI_h/TI_{h-1}T$ we call $T[h]$.
Quotienting further by (\ref{quotientsunday}) (and by (\ref{crossclusters})with
$z=0$ in
the case $h=1$) we obtain ${\cal T}_n\left(\begin{array}{c}
\alpha_h\\
h
\end{array}\right)$.

\begin{pr} The dimension of ${\cal T}_n\left(\begin{array}{c}
\alpha_h\\
h
\end{array}\right)$ is $C_{2n}^{n-h}$.
\end{pr}

Proof. Recall that this number is the number of walks of length $2n$
on the Pascal triangle
from the beginning to height $2h$.
 We associate $U_{1.}\ldots U_{(n-h).}$ to the
lowest such walk that does not visit negative heights. The rest of the
correspondence is obtained by adding diamonds
$\Huge{\Diamond}\!\!\!\!^{\cup}_{\cap}$
 to represent each $U$ until
we reach the same walk again via spacial periodicity
and relation (\ref{quotientsunday}).
 Note that any periodic rectangle
defines a different but isomorphic basis by the relation
(\ref{quotientsunday}).

The intertwiner given in the case $TI_0T$ is also an intertwiner between
${\cal T}_n\left(\begin{array}{c}
\alpha_h\\
h
\end{array}\right)$ and $R_n\left(\begin{array}{c}
x\\
h
\end{array}\right)$  where $\alpha_h=x^h$.

The commutative diagram  (\ref{comdiag}) holds as well here, allowing us to
identify a sequence of invariant subspaces $\rho_n\left(\begin{array}{c}
k\\
h
\end{array}\right)$ of dimension $C_{2n}^{n-h}-C_{2n}^{n-k}$, $h<k\leq n$. By a
similar argument to the case $h=0$ we can find an upper bound on the $x$-order
 of
the Gram matrix $\sum_{k=h+1}^n C_{2n}^{n-k}$. The above sequence of invariant
subspaces provides enough power of $\alpha_h$ to saturate this bound.
The determinant
of the Gram matrix reads therefore
\begin{equation}
\hbox{det
Gr}_{n,h}=\pm\prod_{i=h+1}^{n}\left(-(x+x^{-1})+P_{2i}\right)^{C_{2n}^{n-i}}
\end{equation}
and
\begin{pr}
The representation $R_n\left(\begin{array}{c}
x\\
h
\end{array}\right)$ has dimension $C_{2n}^{n-h}$ and
is irreducible except at $x=q^{\pm 2k}$ $k=h+1,\ldots, n$
. When $x=q^{\pm 2k}$ it contains  $\rho_n\left(\begin{array}{c}
k\\
h
\end{array}\right)$ with dimension $C_{2n}^{n-h}-C_{2n}^{n-k}$
as an irreducible component.
\end{pr}

The representation $\rho_n\left(\begin{array}{c}
h+1\\
h
\end{array}\right)$ is the one isomorphic to the representation
 induced from $C_{n,h}$
by braid translation.

\subsection{The Potts model}\label{The Potts model}

As an application we now determine which of the above irreducible
 representations appear
in the toroidal Potts model. Because of its relationship with the dichromatic
polynomial we know that this is $Q$ generically a direct sum of
$D_{\hat{A}_n}(Q)$ representations (ie $\hat{C}_{n,i}\times\Z_i$), but this
does not tell us the multiplicities. These are helpful in determining the
$Q$-exceptional structure. Consider therefore a most general inhomogeneous
$Q$ state Potts model on  $\hat{A}_{n}\times\Z$ with periodic boundary
conditions in time direction.
Couplings can vary from edge to edge, corresponding to
inhomogeneous edge transfer matrices of the form given in the introduction.
 We restrict for simplicity
to $Q\in[0,4]$ and for $Q$ not an integer define the Potts model
 partition function by the
usual dichromatic polynomial.
Introduce also the partition function $Z_{h,h'}$ of the associated
 6-vertex model with periodic boundary
conditions and such that the total $H$ number  encountered
along the space (resp. time) direction  is equal to $h$ ($h'$). Consider the
same model with twisted boundary conditions in space direction, as given in
\cite{PS}. Then the trace of the transfer matrix in the representation
$\left(\begin{array}{c}
x\\
h
\end{array}\right)$ reads
\begin{equation}
Z\left(\begin{array}{c}
x\\
h
\end{array}\right)=\sum_{h'}x^{h'}Z_{h,h'}\label{fourier}
\end{equation}
In this formula and in what follows we do not take into account the truncation
of various sums due to the finite size of the system, which can be easily
reinstalled. Let us set
\begin{equation}
q=\mbox{exp}\left(\frac{i\pi}{{\cal N}}\right)
\end{equation}
where ${\cal N}$ is a real number. It will turn out that we can restrict to the
case where $x$ is of modulus one, which we can always parametrize as
 $x=q^{2t}$, $t\in[0,{\cal N}]$.
. Then (\ref{fourier}) can be inverted to give
\begin{equation}
Z_{h,h'}=\frac{1}{{\cal N}}\int_{0}^{{\cal
N}}Z\left(\begin{array}{c}
x=q^{2t}\\
h
\end{array}\right)e^{-2i\pi th'/{\cal N}}d\ t\label{fourierinverse}
\end{equation}
The partition function of the $Q$ state Potts model was then rewritten in
\cite{DFSZ} as
\begin{equation}
Z=Z^{(1)}+Z^{(2)}
\end{equation}
with
\begin{equation}
Z^{(1)}=\sum_{h,h'}Z_{h,h'}\mbox{cos}\left[\frac{2\pi}{{\cal
N}}(h\wedge h')\right]
\end{equation}
and
\begin{equation}
Z^{(2)}=\frac{Q-1}{2}
\sum_{h,h'}Z_{h,h'}\mbox{cos}\left[\pi
(h\wedge h')\right]
\end{equation}
where $\wedge$ denotes the  greatest common divisor. This  was
 established by
 analysis of the weights of non contractible loops (in space and time
directions) in the vertex-boundaries correspondence.
To get the desired result we therefore have to plug in these formulas the
inverse Fourier transform (\ref{fourierinverse}). Let us start by $Z^{(2)}$.
Breaking the sum over $h$ into odd and even values gives immediately
\begin{equation}
Z^{(2)}=\frac{Q-1}{2}\left\{\sum_{h\hbox{\tiny{even}}
\neq 0}2Z\left(\begin{array}{c}
x=-1\\
h
\end{array}\right)-Z\left(\begin{array}{c}
x=1\\
h
\end{array}\right)-\sum_{h\hbox{\tiny{odd}}}Z\left(\begin{array}{c}
x=1\\
h
\end{array}\right)+Z\left(\begin{array}{c}
x=-1\\
0
\end{array}\right)\right\}
\end{equation}
As far as $Z^{(1)}$ is concerned, its analysis is more difficult since one has
to take into account the whole arithmetic properties of the spins $h,(h)'$.
For integers $p_1,\ldots,p_k,n_1,\ldots,n_k$ introduce  the expression
\begin{equation}
\left<\hbox{cos}\frac{2\pi}{{\cal N}}p_1^{n_1}\ldots p_k^{n_k}\right>_c
=\sum_{0\leq
m_i\leq\hbox{\tiny{inf}}(n_i,1)}(-)^{\sum m_i}\hbox{cos}\left(\frac{2\pi}{{\cal
N}}p_1^{n_1-m_1}\ldots p_k^{n_k-m_k}\right)
\end{equation}
then the contribution of a given sector $h\neq 0$ is seen to be
\[
\sum_{{\cal M}|h,{\cal M}=p_1^{n_1}\ldots p_k^{n_k}}
\left<\hbox{cos}\frac{2\pi}{{\cal N}}p_1^{n_1}\ldots p_k^{n_k}\right>_c
\left\{ Z\left(\begin{array}{c}
x=1\\
h
\end{array}\right)\right.\hspace{1.5in}
\]
\begin{equation}
\left.+Z\left(\begin{array}{c}
x=e^{2i\pi/{\cal M}}\\
h
\end{array}\right)+Z\left(\begin{array}{c}
x=e^{4i\pi/{\cal M}}\\
h
\end{array}\right)+\ldots+Z\left(\begin{array}{c}
x=e^{2i\pi ({\cal M}-1)/{\cal M}}\\
h
\end{array}\right)\right\}
\end{equation}
It thus decomposes onto  generically irreducible representations.
{}From $h=0$ one just gets $Z\left(\begin{array}{c}
x=q^2\\
0
\end{array}\right)$, which is trace over a reducible representation. However
this trace has to be combined with a trace over $Z\left(\begin{array}{c}
x=1\\
h=1
\end{array}\right)$ that comes with multiplicity $Q-2$ in $Z^{(1)}$ and
$-(Q-1)$ in $Z^{(2)}$ (we used the symmetry $h\rightarrow -h$).
 So on the whole we get trace over the irreducible
 $\rho\left(\begin{array}{c}
1\\
0
\end{array}\right)$.

We therefore conclude that the representations of $T_{\hat{A}_n}(Q)$ appearing
in the Potts model case are  $\rho\left(\begin{array}{c}
1\\
0
\end{array}\right)=\hat{C}_{n,0}$ (with multiplicity one),
 $\rho'\left(\begin{array}{c}
x=-1\\
0
\end{array}\right)=\hat{C}_{n,1}$ (with multiplicity $Q-1$, as expected for the
order parameter), $R\left(\begin{array}{c}
x=e^{2i\pi k/{\cal M}}\\
h
\end{array}\right)$ for $k=1,\ldots,{\cal M}-1$ and ${\cal M}|h$, $h$ arbitrary
(but limited for a given $n$). This is in
complete agreement with the above discussion of the $D_{\hat{A}_n}$ case.

Let us now discuss multiplicites. For
the case $h=2$ we get $R\left(\begin{array}{c}
x=1\\
h=2
\end{array}\right),R\left(\begin{array}{c}
x=-1\\
h=2
\end{array}\right)$. The first one comes with multiplicity $Q^2-4Q+2$ in
$Z^{(1)}$ and $-Q+1$ in $Z^{(2)}$, adding to $Q^2-5Q+3$.
 The second comes with $Q^2-5Q+4$ in
$Z^{(1)}$ and $2Q-2$ in $Z^{(2)}$, adding to $Q^2-3Q+2$. In general the
representation $R\left(\begin{array}{c}
x=1\\
h
\end{array}\right)$ arises with multiplicity $P_{2h}-Q+1$.
The representation $R\left(\begin{array}{c}
x=-1\\
h
\end{array}\right)$ that arises only for $h$ even has multiplicity
$P_{2h}-P_2+2Q-2$. All other representations $\left(\begin{array}{c}
x=e^{2i\pi {\cal P}/{\cal Q}}\\
h
\end{array}\right)$ (${\cal P}<{\cal Q}$ coprimes)
 arise with the multiplicity, if we write the decomposition in prime factors
 ${\cal Q}=p_1^{l_1}\ldots
p_k^{l_k}$, $h=p_1^{n_1}\ldots p_k^{n_k}$
\begin{equation}
\sum_{l_i\leq m_i\leq n_i}2\left<\hbox{cos}\frac{2\pi}{{\cal N}}p_1^{m_1}\ldots
p_k^{m_k}\right>_c
\end{equation}
These can be rewritten as polynomials in $Q$, with no very illuminating forms
however.  The multiplicity of $R_n\left(\begin{array}{c}
x\\
h
\end{array}\right)$ is independent of $n$ for $n$ large enough, as in the $A_n$
and mean field case. In those cases the result is a consequence of theorem
\ref{fet}. Here this theorem does not hold.

\section{Arbitrary Graphs and the Full Case}

\subsection{Examples}

Let us discuss a few examples. Consider first the Daisy graph $G=D_4$. As usual
we start with $I_0$ and generate $TI_0$ by acting with the generators. Choose
the convention that the central node in such a Daisy graph is associated
with $U_{n.}$ for $n-1$
legs. Then list the words
\begin{equation}
\begin{array}{cccccccc}
I_0\\
U_{14}I_0&U_{24}I_0&U_{34}I_0\\
U_{14}U_{24}I_0&U_{14}U_{34}I_0&U_{24}U_{34}I_0\\
U_{4.}U_{14}U_{24}I_0&U_{4.}U_{14}U_{34}I_0&U_{4.}U_{24}U_{34}I_0&
U_{14}U_{24}U_{34}I_0\\
U_{34}U_{4.}U_{14}U_{24}I_0&U_{24}U_{4.}U_{14}U_{34}I_0
&U_{14}U_{4.}U_{24}U_{34}I_0&
U_{4.}U_{14}U_{24}U_{34}I_0\\
\end{array}\label{garbageI}
\end{equation}
The basis truncates at this point because the relations collapse all longer
words onto these. Therefore the ideal $TI_0$ is finite dimensional (and
dim($TI_0$)=dim($D_GI_0$)=dim($D_{\underline{n}}I_0$)).

Consider now the graph $G=D_5^{(1)}$ with the incidence matrix
\begin{equation}
\left(\begin{array}{cccccc}
0&1&0&0&0&0\\
1&0&1&0&1&0\\
0&1&0&0&0&0\\
0&0&0&0&1&0\\
0&1&0&1&0&1\\
0&0&0&0&1&0
\end{array}\right)
\end{equation}
Rather than generate the whole basis, we simply note that the following set of
words does not collapse
\begin{equation}
\left(I_0U_{23}U_{25}U_{45}U_{2.}U_{5.}U_{12}U_{25}U_{56}U_{2.}
U_{5.}U_{23}U_{25}U_{45}U_{2.}U_{5.}U_{12}U_{25}U_{56}I_0\right)^k
\end{equation}
so the algebra is infinite dimensional. How can we characterize such non
collapsing words? If we draw the connectivity diagram of the above word on
$G\times\Z$ ,
the interior of each bracket is made of two interlocked loops: a Hopf link. In
the diagram algebra, these would collapse to give two factors of $Q$. But in
the Full algebra case there is no relation to simplify such a word. Recall that
in the ${\hat A}$ case similar words that could not be collapsed corresponded
to non contractible loops.

Returning to $G=D_3$ consider now the $h=1$ sector.
\begin{equation}
{\small\begin{array}{cccccccccccccc}
I_1
\\
\\
U_{14}I_1&U_{24}I_1&U_{34}I_1
\\
\\
U_{14}U_{24}I_1\hspace{.7cm}U_{14}U_{34}I_1&U_{24}U_{34}I_1\hspace{.7cm}
U_{4.}U_{24}I_1&U_{4.}U_{34}I_1\hspace{.7cm}U_{4.}U_{14}I_1
\\
\\
U_{4.}U_{14}U_{24}I_1\hspace{.4cm}U_{4.}U_{14}U_{34}I_1&
U_{4.}U_{24}U_{34}I_1\hspace{.4cm}
U_{14}U_{24}U_{34}I_1&
U_{14}U_{4.}U_{24}I_1\\
U_{24}U_{4.}U_{34}I_1\hspace{.4cm}U_{34}U_{4.}U_{14}I_1&
U_{34}U_{4.}U_{24}I_1\hspace{.4cm}
U_{14}U_{4.}U_{34}I_1&U_{24}U_{4.}U_{14}I_1
\\
\\
U_{14}U_{34}U_{4.}U_{24}I_1&U_{14}U_{24}U_{4.}U_{34}I_1&U_{24}U_{34}
U_{4.}U_{14}I_1\\
U_{34}U_{4.}U_{14}U_{24}I_1&
U_{24}U_{4.}U_{14}U_{34}I_1
&U_{14}U_{4.}U_{24}U_{34}I_1\\
U_{4.}U_{14}U_{24}U_{34}I_1
\\
\\
U_{4.}U_{14}U_{34}U_{4.}U_{24}I_1&U_{4.}U_{14}U_{24}U_{4.}U_{34}I_1&
U_{4.}U_{24}U_{34}U_{4.}U_{14}I_1
\\
\\
U_{24}U_{4.}U_{14}U_{34}U_{4.}U_{24}I_1&U_{34}U_{4.}U_{14}U_{24}U_{4.}U_{34}I_1
&U_{14}U_{4.}U_{24}U_{34}U_{4.}U_{14}I_1
\\
\\
U_{2.}U_{24}U_{4.}U_{14}U_{34}U_{4.}U_{24}I_1&
U_{2.}U_{34}U_{4.}U_{14}U_{24}U_{4.}U_{34}I_1
&U_{2.}U_{14}U_{4.}U_{24}U_{34}U_{4.}U_{14}I_1
\end{array}}
\end{equation}
As in the list in (\ref{garbageI}) the table truncates here.
The total dimension is 36 (cf the $\underline{4}$ case, the state
$\left((123)(4)'\right)$ cannot be reached from $I_1\sim
\left((1)(2)(3)(4)'\right)$). However, as we will see in the next section the
$h=2$ sector is infinite dimensional.

\subsection{On the characterization of irreducible representations.}

Consider the $h=0$ sector. We may associate words in the ideal $TI_0$ with
certain elements of the universal set of $G\times\Z$ (as we do in $D_G$, but
here the equivalence relations are different). Words are
equivalent if their graphs drawn on $G\times\Z$
are related by the following moves

1: contraction of cul-de-sacs and removal of points, up to factors of $Q$

2: deformations of the graph which take a segment to a segment without touching
another segment.

Note that such a deformation may take an empty disconnected loop to a line and
hence remove it at cost a factor of $Q$. To establish these moves note that
objects   such as cul-de-sacs and loops may be embedded in $A_n\times\Z$, where
the moves are allowed by the relations.

The quotient relations required to make the ideal $TI_0$ finite dimensional
must tell us how to deal with each type of object drawn on $G\times\Z$ distinct
under the relations above. Suppose such an object $K$ is not equivalent to the
empty graph. Then formally $(KI_0)^k$ gives a new element of the ideal for each
higher integer $k$. On the other hand $TI_0(KI_0)^k$ is an invariant subspace
of $TI_0$ (c.f. the ${\hat A}_n$ case), so the way to deal with this is to
 introduce quotient relation
\begin{equation}
I_0KI_0=\alpha_K Q^{n(K)} I_0
\end{equation}
where $\alpha_K\in\C$ and $n(K)$ is the number of components in $K$. There is
no constraint on $\alpha_K$ forced by the relations, since $I_0$ forms an
impenetrable barrier (c.f. the relations and the cases $A_n$ and ${\hat A}_n$).

Relations of this kind are not in general sufficient to produce finite
dimensional ideals. However the additional relations required to make a finite
dimensional ideal do not seem to give rise to continuous parameters. Such
additional relations generate still further relations which constrain the value
of any parameter introduced.

The simplest example of both these features is provided by $D_4^{(1)}$. The
normal 52 elements basis of $D_GI_0$ is extended by the non collapse of
words like
\begin{equation}
\left(U_{5.}U_{35}U_{15}U_{5.}U_{25}U_{45}\right)^kI_0
\end{equation}
Accordingly the relations
\begin{equation}
I_0\; U_{35}U_{15}U_{5.}U_{25}U_{45}
\left(U_{5.}U_{35}U_{15}U_{5.}U_{25}U_{45}\right)I_0=
\alpha_KQ^2I_0\label{fivefifteen}
\end{equation}
which can be represented graphically as
\begin{equation}
I_0(\hbox{Hopf Link})I_0=\alpha_K I_0\left(
\begin{picture}(30,0)(0,0)
\put(5,3){\circle{10}}\put(17,3){\circle{10}}
\end{picture}
\!\!\right)I_0
\end{equation}
may be introduced with no restriction on $\alpha_K$.  In the other
 quotient required to get a finite dimensional ideal (and consistent with $D_G$
- i.e. taking connectivity to identical connectivity)
\begin{equation}
\left(U_{5.}U_{35}U_{15}U_{5.}U_{25}U_{45}\right)^{k+2}I_0=z
\left(U_{5.}U_{35}U_{15}U_{5.}U_{25}U_{45}\right)^2I_0
\end{equation}
we are forced to take $z=Q^k$ for consistency with the action of $U_{35}U_{15}
U_{25}U_{45}$. The $D_GI_0$ quotient is the case $k=-1$, for which
$\alpha_K$ is
also determined by the relations.

The smallest representation which supports the quotient (\ref{fivefifteen})
(together with its images under $\Z_4$ which imbed the Hopf link into
$G\times\Z$ in ways that are different with respect to the moves)
has dimension 301. This is the case $k=1$.

Let us give a more interesting example. We will construct a free parameter
quotient associated to the trefoil knot for $G$ the Daisy with 6 petals.
\[
I_0U_{67}U_{57}U_{7.}U_{17}U_{47}U_{7.}U_{67}U_{27}U_{7.}U_{37}U_{17}U_{6.}
U_{7.}U_{67}U_{27}U_{7.}U_{1.}U_{17}U_{37}U_{7.}U_{67}U_{47}U_{7.}U_{17}
U_{57}I_0
\]
\[
\hspace{2in}=\alpha_{\hbox{tref}}\ QI_0
\]
The best way to visualize this is to construct a piece of paper with section
$G$ and imbed the knot in this "oregamoid" such that each sheet carries at most
one piece of string at a time. Incidentally the graph in figure
\ref{backpaper} will also support the trefoil knot with the long leg carrying
up to three lines at once.
\begin{figure}
\vspace{3cm}\caption{A graph supporting the trefoil knot}\label{backpaper}
\end{figure}

These examples can be generalised to give
\begin{pr}
There are at least as many free parameters associated to the ideal $T_GI_0$
as links embeddable in $G\times\Z$.
\end{pr}

In physical terms, connectivities carry information. For instance the case
$h=1$ describes spin spin correlations. The $h=0$ sector may be associated to
the partition function. When we consider the $h>0$ sectors therefore we must
keep track of information at the boundary of the system and hence in the
present language move from links to tangles. For example on the Daisy with 5
petals the `Staffordshire Knot' (the tangle generated by opening one arc of the
trefoil) gives
\begin{eqnarray*}
I_1U_{56}U_{6.}U_{16}U_{46}U_{6.}U_{56}U_{26}U_{6.}U_{36}U_{16}U_{5.}U_{6.}
U_{56}U_{26}U_{6.}U_{1.}U_{16}U_{36}U_{6.}U_{56}U_{46}U_{6.}U_{16}I_1\\
\hspace{1.5in}=
\alpha_{\hbox{Staff}}\ QI_1\ \left(\hbox{mod}TI_0T\right).
\end{eqnarray*}

Since  for any $G$ other than $A_n$ two strands can
be braided twice on $G\times\Z$ we have

\begin{pr}
All algebras $T_G\ $ ($G\neq A_n$) are infinite dimensional.
\end{pr}

For instance for $G=D_3$  (see figure \ref{backpaperII})
\begin{equation}
b=I_2U_{41}U_{4.}U_{43}U_{42}U_{3.}U_{4.}U_{43}U_{41}U_{4.}U_{42}I_2
\end{equation}
is braided once, so $b^2=\alpha I_2$ is a legitimate quotient.
\begin{figure}
\vspace{5cm}\caption{A braid can be drawn on $D_4\times\Z$}\label{backpaperII}
\end{figure}

\section{Remarks}

\subsection{$Q$ non generic}

Until now we considered only exceptional values of the free parameters. Let us
now consider exceptional values of $Q$. If there are free parameters, this is
only interesting at their exceptional values (see e.g. $\hbox{Gr}_{2,0}$).

This is has been understood in details for $A_n$ \cite{Mar,HGJ}. The
non generic values in that case are given by
\begin{equation}
Q=4\left(\hbox{cos}\frac{\pi p}{q}\right)^2\label{nongeneric}
\end{equation}
where $p,q$ are coprime integers. The case $p=1$ is usually referred to as
Beraha numbers. The continuum limit of the corresponding
ABF \cite{ABF} models is given by a `minimal' conformal field theory \cite{BPZ}
which is unitary in the $p=1$ case.

The non generic $Q$ values in the ${\hat A}_n$ case are identical, as can be
expected since the difference with $A_n$ is merely a `boundary effect'.

In the mean field case preliminary study shows that at least all integer $Q$
are non generic, and it seems likely that there are no other.

In the $A_n,{\hat A}_n$ cases, the maximum non generic value of $Q$
(\ref{nongeneric}) is $Q=4$, which
 coincides with the boundary between second and
first oder phase transitions. As we `increase the dimension' to the mean field
case $G=\underline{n}$ these two numbers move therefore in opposite directions.
Indeed in the mean field case the boundary between second and first order phase
transtions is now at $Q=2$ while, as mentioned above, the set of non generic
$Q$ is unbounded.

A more complete study of the non generic $Q$ case for ${\hat A}_n$ and
$\underline{n}$ will appear elsewhere.

In two dimensions, relations between $Q$ non generic values and colouring
problems have been observed \cite{???}. These relations
seem to extend to the mean field
case.

\subsection{The General $D_{Daisy}$}

The dimensions of the irreducible representations for the general Daisy case
can be computed. The numbers themshelves are not very illuminating. However we
note that $D_GI_0$ has the same dimension as the complete graph  case
(with same
number of vertices). Therefore  the asymptotic rate of growth of
dimensions is unbounded.

\subsection{Conclusion}

Despite the difficulty of the higher dimensional case, such as $G=A_n\times
A_n$, the study of algebraic properties seems an interesting alternative to the
search  for integrable systems. Among the important questions amenable with
existing techniques (or computers) are the determination of dimensions of
generically irreducible representations, and hence  exceptional $Q$ values.
A warm up case would be $G=A_n\times A_2$.

\bigskip

{\bf Acknowledgements}

P.P.M would like to thank
the UK SERC, City University (London)
 and the Nuffield Foundation for support, and
 B.Westbury for useful conversations. This work was supported by the Packard
Foundation and DOE Contract DE-AC02-76ERO3075.

\end{document}